\documentclass[twocolumn]{aastex631}

\usepackage{times}
\usepackage{amsmath,amssymb,amstext}
\hypersetup{linkcolor=red,citecolor=blue,filecolor=cyan,urlcolor=blue}
\usepackage{tabularx}
\usepackage{longtable}
\usepackage{float}
\usepackage{natbib}
\usepackage{graphicx,color}
\usepackage{txfonts}

\usepackage{multirow}
\usepackage{array}
\usepackage{rotating}
\usepackage{sidecap}
\usepackage{hyperref}
\usepackage{epstopdf}
\usepackage{footnote}
\usepackage{tabularx}
\usepackage{booktabs}
\usepackage{longtable}
\usepackage{tabu}
\usepackage{longtable}

\newcommand{\kms}{km\,s$^{-1}$}

\begin{document}

\title{{\large{\bf Quasiperiodic Energy Release and Jets at the Base of Solar Coronal Plumes}}}
\author{Pankaj Kumar\altaffiliation{1,2}}

\affiliation{Department of Physics, American University, Washington, DC 20016, USA}
\affiliation{Heliophysics Science Division, NASA Goddard Space Flight Center, Greenbelt, MD, 20771, USA}

\author{Judith T.\ Karpen}
\affiliation{Heliophysics Science Division, NASA Goddard Space Flight Center, Greenbelt, MD, 20771, USA}

\author{Vadim M. Uritsky\altaffiliation{3,1}}
\affiliation{Catholic University of America, 620 Michigan Avenue NE, Washington, DC 20064, USA}
\affiliation{Heliophysics Science Division, NASA Goddard Space Flight Center, Greenbelt, MD, 20771, USA}

\author{Craig E. Deforest}
\affiliation{Southwest Research Institute, 1050 Walnut Street, Suite 300, Boulder, CO 80302, USA}

\author{Nour E. Raouafi}
\affiliation{The John Hopkins University Applied Physics Laboratory, Laurel, MD 20723, USA}

\author{C.\ Richard DeVore}
\affiliation{Heliophysics Science Division, NASA Goddard Space Flight Center, Greenbelt, MD, 20771, USA}

\email{pankaj.kumar@nasa.gov}

\begin{abstract}
 Coronal plumes are long, ray-like, open structures, which have been considered as possible sources for the solar wind. 
Their origin in the largely unipolar coronal holes has long been a mystery. Earlier spectroscopic and imaging observations revealed blue-shifted plasma and propagating disturbances (PDs) in plumes that are widely interpreted in terms of flows and/or propagating slow-mode waves, but these interpretations (flows vs waves) remain under debate. Recently we discovered an important clue about plume internal structure: dynamic filamentary features called ``plumelets'', which account for most of the plume emission.   Here we present high-resolution observations from the Solar Dynamics Observatory’s Atmospheric Imaging Assembly (SDO/AIA) and the Interface Region Imaging Spectrograph (IRIS) that revealed numerous, quasiperiodic, tiny jets (so-called ``jetlets") associated with transient brightening, flows, and plasma heating at the chromospheric footpoints of the plumelets. By analogy to larger coronal jets, these jetlets are most likely produced within the plume base by magnetic reconnection between closed and open flux at stressed 3D null points. The jetlet-associated brightenings are in phase with plumelet-associated PDs, and vary with a period of $\sim$3 to 5 minutes, which is remarkably consistent with the photospheric/chromospheric p-mode oscillation. This reconnection at the open-closed boundary in the chromosphere/transition region is likely modulated or driven by local manifestations of the global p-mode waves. The jetlets extend upward to become plumelets, contribute mass to the solar wind, and may be sources of the switchbacks recently detected by the Parker Solar Probe.
\end{abstract}
\keywords{Sun: jets---Sun: corona---Sun: UV radiation---Sun: magnetic fields---Sun: coronal holes}


\section{INTRODUCTION}\label{intro}
Plumes are persistent, columnar structures embedded in coronal holes (CHs), detectable in visible light as ``coronal rays" \citep{saito1958} and in ultraviolet/extreme-ultraviolet (UV/EUV) emissions. They have been observed for as long as the corona has been marveled at during solar eclipses, and measured extensively for over 100 years \citep{bigelow1891}. Plumes expand super-radially from photospheric flux concentrations (plages) into the outer corona \citep{fisher1995,deforest2001a,raouafi2007}, and have been detected out to $\sim$40 R${_\odot}$ \citep{woo1997,deforest2001b}. 
While each plume brightens and stands out in EUV movies for only a few hours to a day, they recur at the same location many times during a lifetime of several weeks \citep{deforest2001b}. Although plumes appear in unprocessed images as diffuse, columnar features that taper to smaller footprints at the base, we have found that all plume emission comes from bright filamentary substructures denoted ``plumelets" \citep{uritsky2021}.
Individual plumelets fluctuate in intensity on time scales of a few minutes, far faster than the hours-long variations of the bulk plume brightness  \citep{deforest1997,deforest2007,raouafi2014,uritsky2021}.  Analysis of STEREO/EUVI images revealed the presence of quasiperiodic fluctuations in plume brightness, denoted propagating disturbances (PDs), with $\sim$5-30 min periods \citep{mcintosh2010,tian2011}; these perturbations were attributed to jets adding mass to the plumes. In contrast, other studies of SOHO/EIT and SDO/AIA plume observations have interpreted such PDs as evidence of slow magnetosonic waves \citep{deforest1998,ofman1999,banerjee2016,banerjee2021}.  \citet{gupta2012} reported the unique detection of PDs with a period of 14.5 min in a polar plume, based on SOHO/SUMER spectroscopic observations. Although such spectroscopic observations of PDs are important for identifying their true physical nature, they are rare and have not addressed variability at the plume base. The controversy on flows vs waves in plumes is still under debate \citep{poletto2015,wang2016}.  Our research on reconnection-driven coronal jets, extended here to smaller scales, suggests that both waves and flows could coexist in plumes.

Additional evidence for small-scale energy release low in the solar atmosphere comes from SOHO/EIT and Hi-C ($\approx$3 minute observation) observations of faint EUV brightening (or bright dots) at the footpoint of active region fan-loops \citep{berghmans1999,regnier2014}. 
Recent SolO observations have revealed similar small EUV ``campfires" rooted in the chromospheric network in a quiet-Sun region, with unprecedented resolution and stereoscopic height estimates made with simultaneous SDO/AIA views \citep{berghmans2021}. None of these events were reported to be associated with outflows or waves, however, and their sources were not located at the base of plumes. 


The advent of high-resolution, high-cadence coronal observations from the Solar Dynamics Observatory's Atmospheric Imaging Assembly (SDO/AIA), coupled with photospheric magnetograms from SDO's Helioseismic and Magnetic Imager (SDO/HMI), has enabled detailed studies of plumes from their footpoints outward.  Prior investigations of plume bases revealed the existence of tiny jets (denoted ``jetlets") above minority-polarity intrusions, accompanied by EUV brightenings in the low corona  \citep{raouafi2008, raouafi2014, pant2015, panesar2018}. The detection of these jetlets at the base of plumes led to the hypothesis that they are the long-sought source of mass and energy that sustains plumes for hours to weeks \citep{raouafi2014}.

Magnetic reconnection in an embedded-bipole topology is broadly accepted as the driver of coronal jets and other eruptive solar phenomena \citep{sterling2015,wyper2017,kumar2018,kumar2019a,kumar2019b}. Therefore the underlying magnetic-field inhomogeneity is fundamental to the existence of jets, and by extension to that of plumes. The bursty, localized nature of reconnection appears at first glance to be incompatible with the observed long lifetime of plumes and with quasisteady wind acceleration \citep{tu2005}. However, studies of impulsive coronal-heating mechanisms (e.g., nanoflares) have demonstrated that myriad small reconnection events in complex field geometries collectively can produce quasisteady heating and flows \citep{klimchuk2006,klimchuk2015}. Observations of numerous small energetic events such as ``jetlets" and ``plume transient bright points'' occurring across the bases of several plumes suggest an analogous scenario for the creation and maintenance of plumes \citep{raouafi2014}. Numerical studies of individual reconnection-driven jets generally support this contention \citep{karpen2017,uritsky2017,roberts2018}, but the effects of collective jet activity have not yet been modeled. 

The magnetic topology, frequency, and energy of the jetlets, as well as their connections with plumelets and their PDs, were not fully explored by previous investigations. Therefore, the present study is focused on answering the following questions. (i) How frequently are jetlets observed at the base of plumes? (ii) What are their magnetic topology and triggering mechanism? (iii) How are jetlet flows associated with other signatures of energy release at the plume base? (iv) What is the total energy of jetlets, and how much mass flux do they contribute to the solar wind? 


We have analyzed high-cadence multiwavelength SDO/AIA and HMI observations for well-observed plumes, focusing on the activity at the base and connections to the fine structure within the overlying plume (``plumelets" already described by  \citet{uritsky2021}). In contrast to earlier studies, we used a noise-gating method \citep{deforest2017} to clean the AIA and HMI data, which revealed in exquisite detail the jetlets and other small-scale structures emanating from the plume bases. For the first time, we observed repeated EUV brightenings and associated jetlets at the bases of plumes (best observed in AIA 193/171 \AA) with a period of $\sim$3-5 min.  Our investigation identified multiple quasiperiodic jetlets within the mixed-polarity plume base, throughout the observation intervals (up to 40 hours), which evolve into the plumelets that comprise the overlying bright plume. We discuss the measured and derived jetlet properties, the structural and dynamic connections between the jetlets and the plumes, and implications for the underlying physical processes. These observations are a critical step towards resolving the long-standing mystery of the origin of plumes as well as the long-lasting flows vs wave controversy. Although our study finds that jetlets are bulk flows, not slow-mode waves, episodic, spatially distributed reconnection events at the base of a plume are likely to generate outward-directed MHD waves as well as outflows. In addition to generating jetlets and plumelets, these small-scale reconnection events are expected to contribute to plume heating and the mass of the nascent solar wind.

\begin{figure*}
\centering{
\includegraphics[width=17cm]{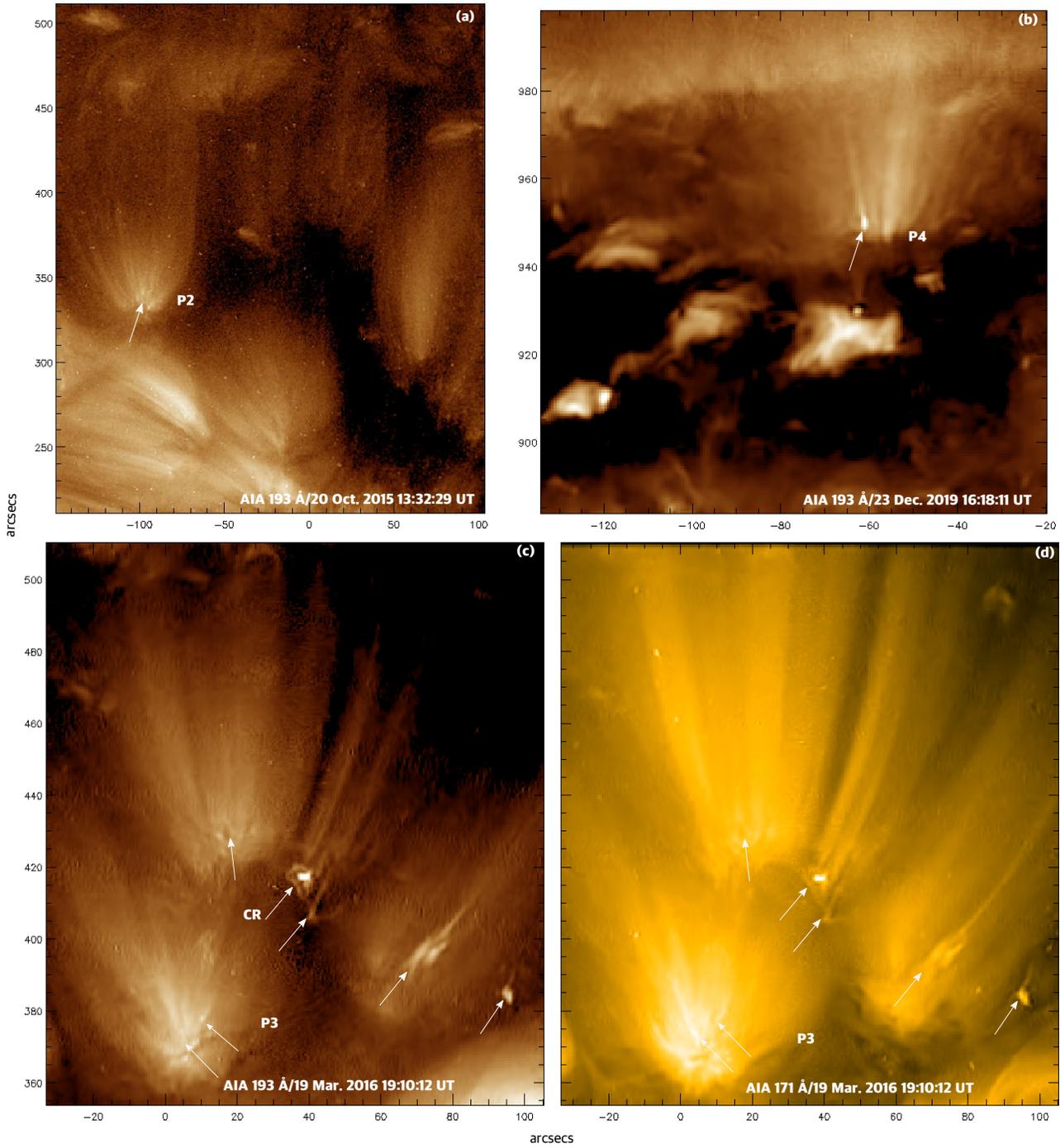}
}
\caption{Plumes P2, P3, and P4 in equatorial and polar CHs. Arrows indicate tiny brightenings associated with jetlets. Plume P3 (c,d) and neighboring plumes were rooted in a curved plage/network region. CR=circular (quasi) ribbon.} 
\label{fig1}
\end{figure*}

\begin{figure*}
\centering{
\includegraphics[width=17cm]{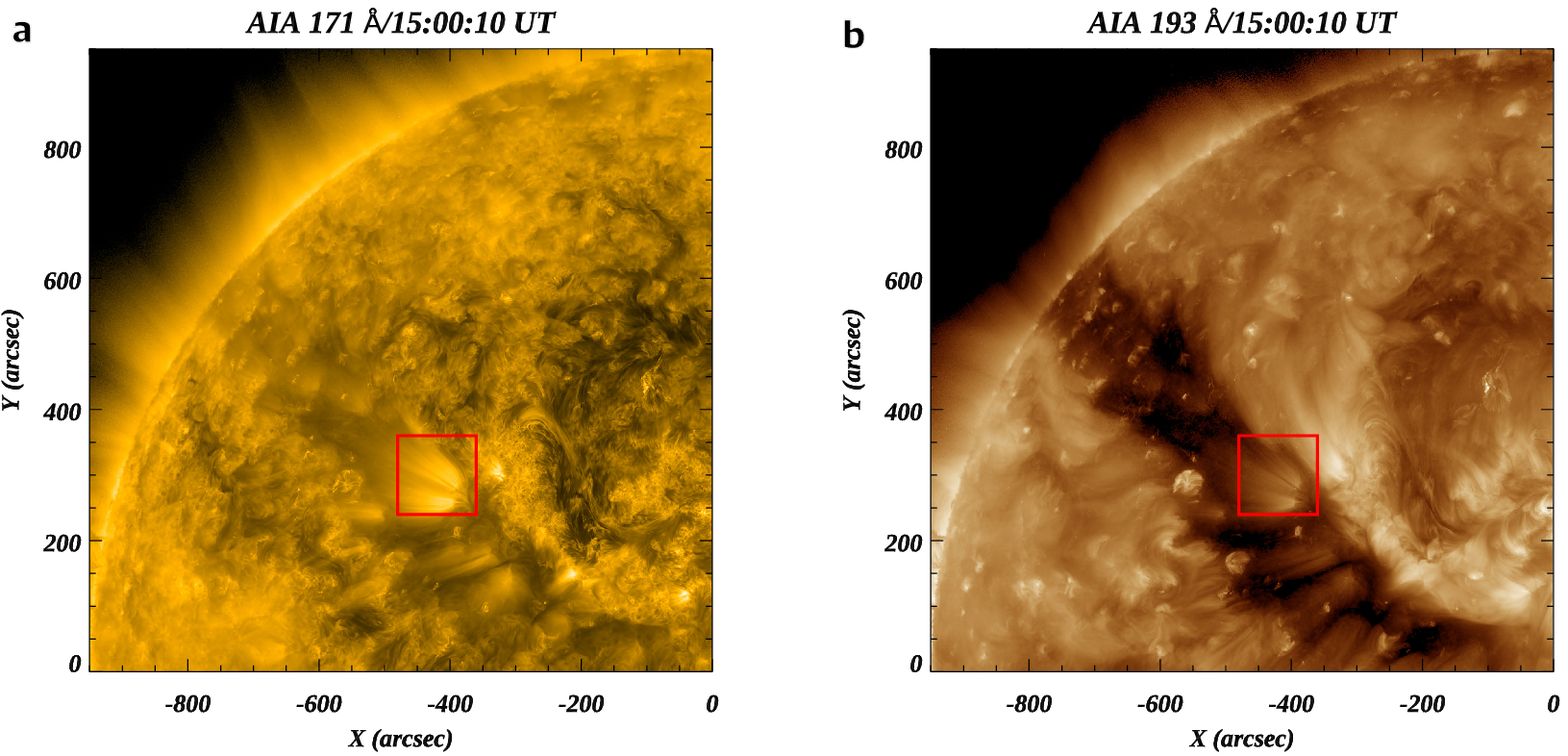}
\includegraphics[width=17cm]{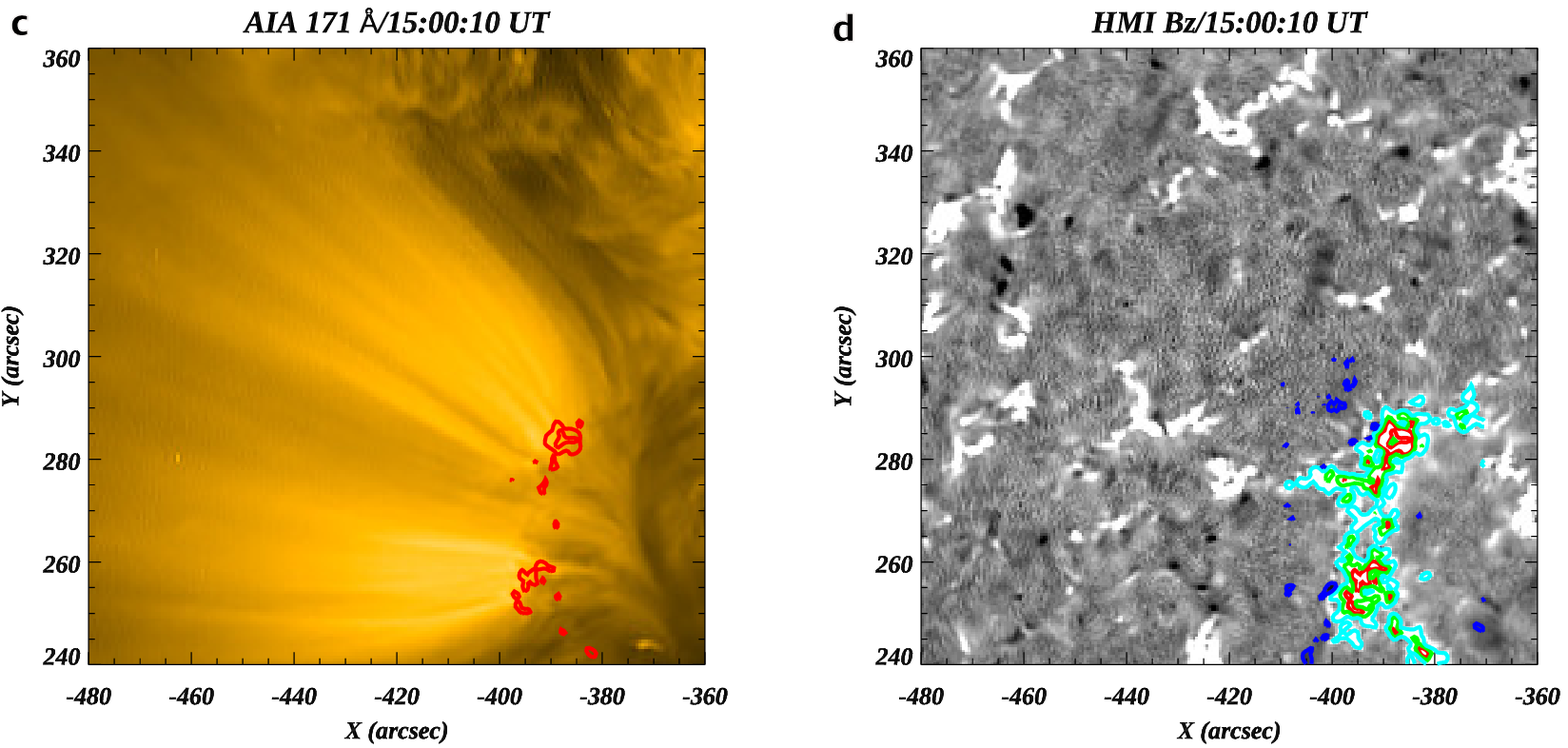}
}
\caption{Location of plume P1 in an equatorial CH at 15:00:10 UT on 3 July 2016. (a,b) AIA 171 and 193 \AA~ images showing plume structure (within red box) in the coronal hole. (c) Zoomed view of the plume in the AIA 171 \AA~ channel overlaid by HMI magnetogram contours of +400 Gauss. (d) HMI line-of-sight magnetogram (scale=$\pm$30 G). The contour levels at the footpoint of the plume are +400 G (red), +200 G (green), and +50 G (cyan). The blue contours (-10 G) are weak opposite (minority) polarities at the boundary of the plume base.} 
\label{fig2}
\end{figure*}
\begin{figure*}
\centering{
\includegraphics[width=17.5cm]{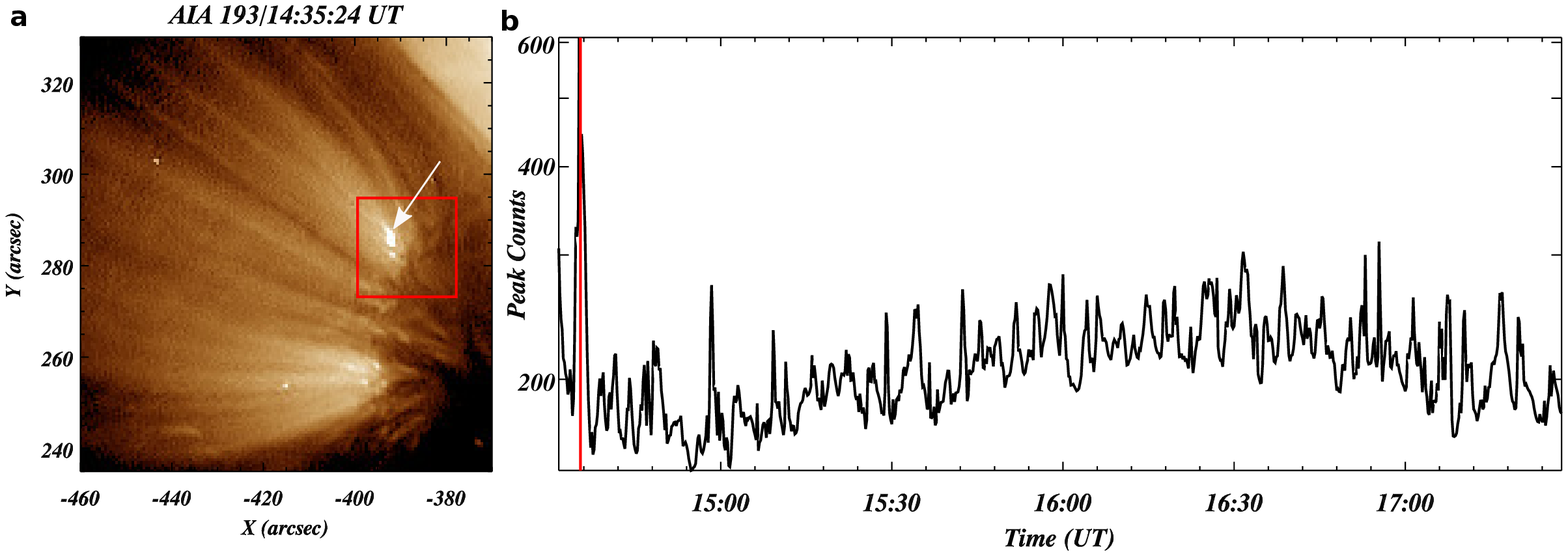}
\includegraphics[width=17.5cm]{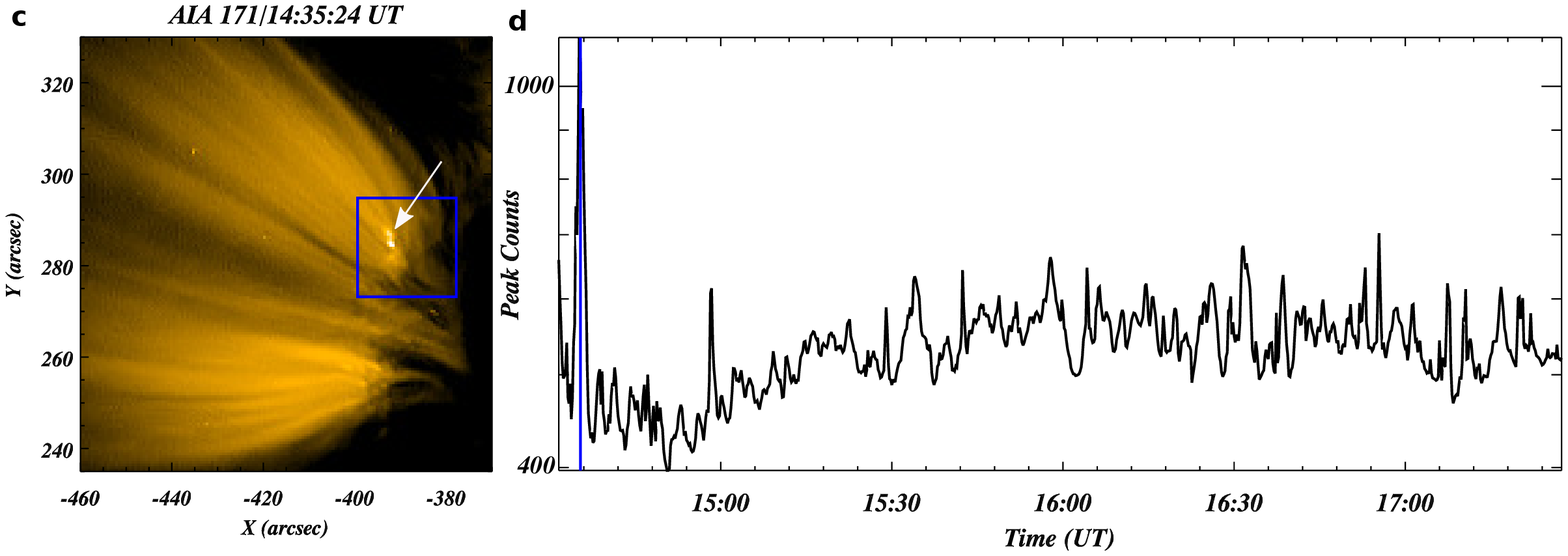}
}
\caption{Brightenings near the base of plume P1 associated with jetlets. (a,c) AIA 193/171 \AA~ image of the plume within an equatorial CH. The white arrow indicates one intense brightening associated with a tiny jet along the plume. (b,d) Temporal evolution of the peak counts (arbitrary units) in two AIA channels, extracted from the red/blue boxes. A vertical red/blue line indicates the timing of the 193/171 \AA~ image. (An animation of this figure is available.)} 
\label{fig3}
\end{figure*}


\begin{figure*}
\centering{
\includegraphics[width=11.0cm]{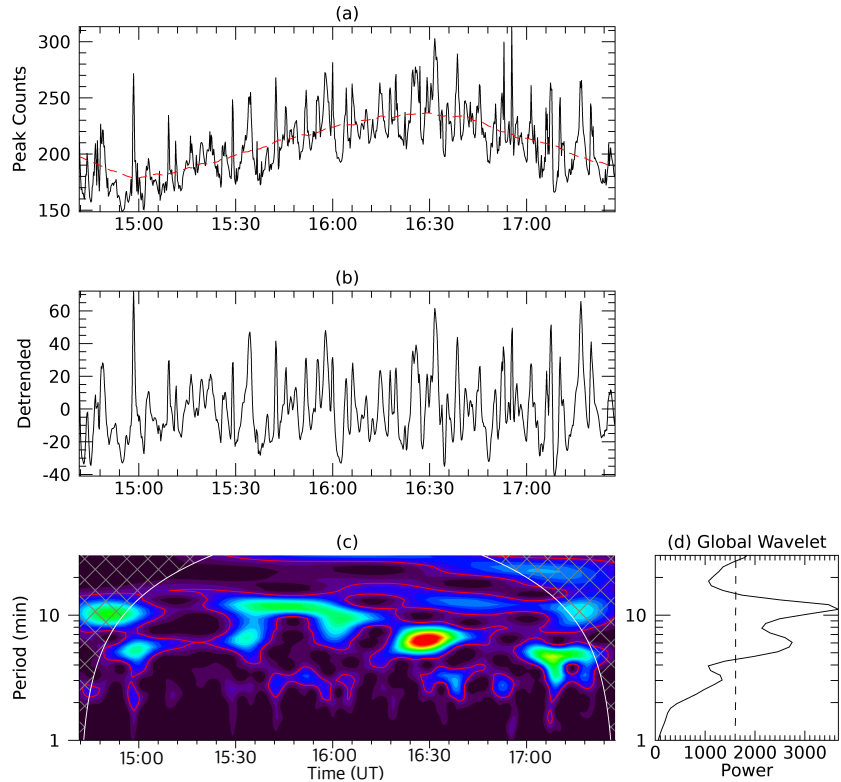}
}
\caption{Periodicity of jetlet-associated brightenings at the base of plume P1: (a) Intensity variations in a selected region of the AIA 193  \AA~ image (same as Fig. \ref{fig3}b). (b) The smoothed and detrended light curve after subtracting the red trend shown in (a) from the original intensity. (c) Wavelet power spectrum of the detrended signal. Red contours outline the 99$\%$ significance level. (d) Global wavelet power spectrum. The dashed line is 99$\%$ global confidence level.} 
\label{fig4}
\end{figure*}

\begin{figure*}
\centering{
\includegraphics[width=17cm]{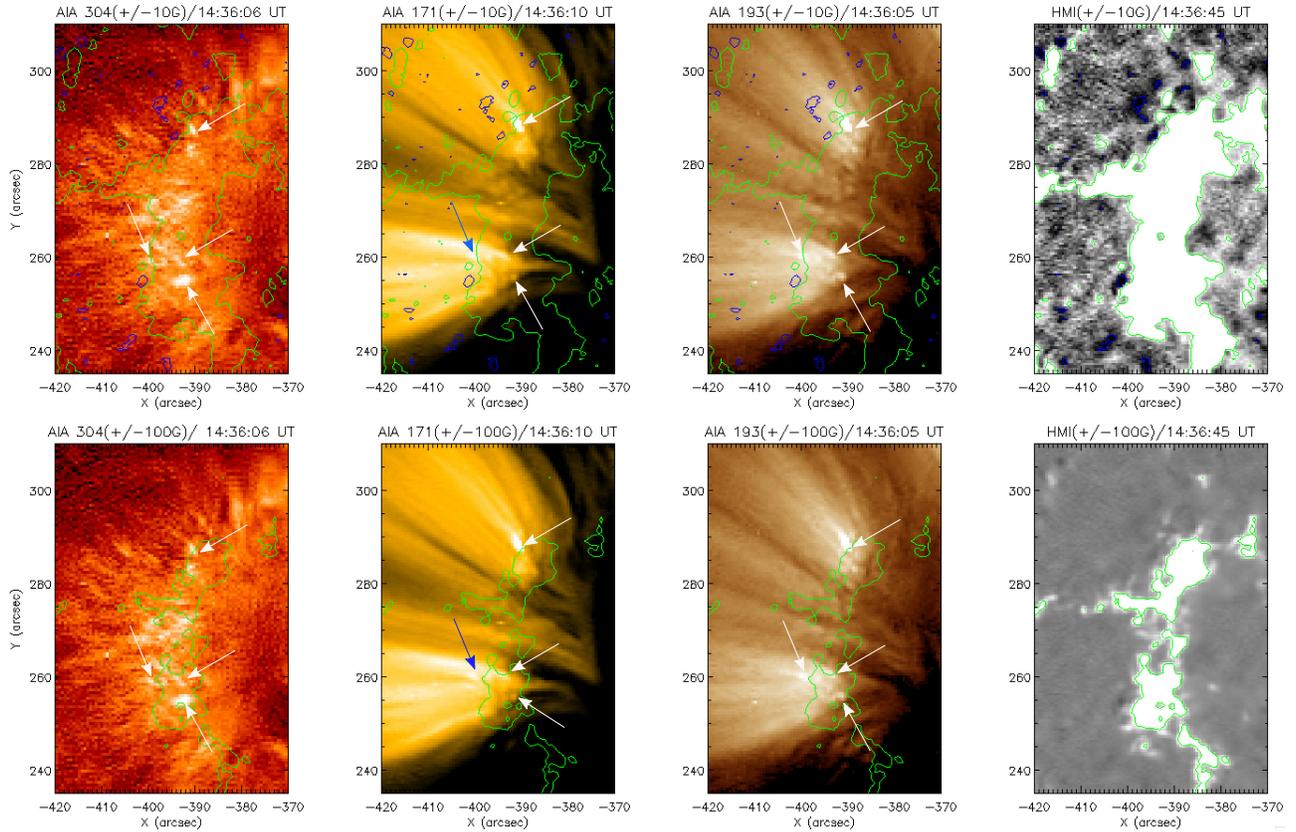}
}
\caption{Left to right: AIA 304, 171, and 193 \AA~ images of jetlets and associated brightenings (marked by arrows), plus a nearly cotemporal HMI magnetogram with photospheric magnetic field strength of $\pm$10 G (top) and $\pm$100 G (bottom) at the base of plume P1 on July 3 2016. The top (bottom) panels are overlaid by HMI magnetogram contours of $\pm$10 ($\pm$100) G. Green (blue) is positive (negative). (An animation of this figure is available.)} 
\label{fig5}
\end{figure*}

\begin{figure*}
\centering{
\includegraphics[width=18cm]{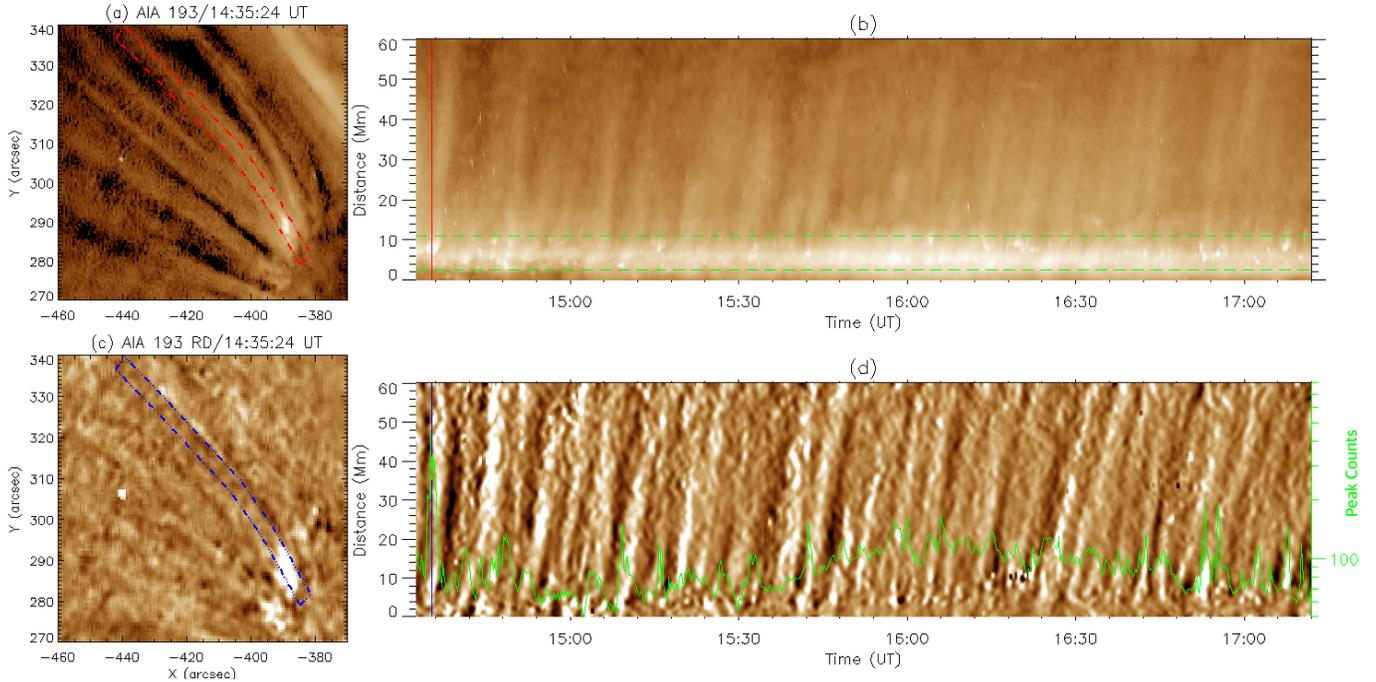}
}
\caption{AIA 193 \AA~ (a) intensity and (c) running-difference images of jetlets and associated brightenings at the base of plume P1 on July 3 2016. Time-distance plots of (b) 193 \AA~ total intensity and (d) 193 \AA~ running-difference intensity along the rectangular slits (red and blue dashed outlines) shown in (a) and (c). The green curve in panel (d) shows peak 193 \AA~ counts extracted from the plume's footpoint region marked between two horizontal green dashed lines in panel (b). (An animation of this figure is available.)} 
\label{fig6}
\end{figure*}

\begin{figure}
\centering{
\includegraphics[width=9cm]{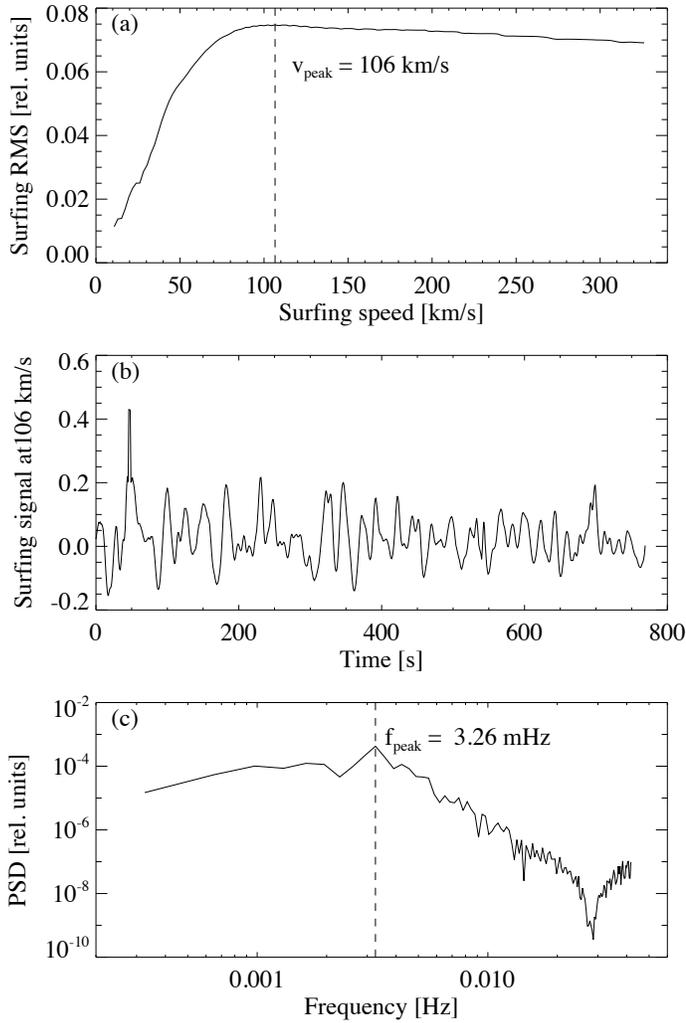}
}
\caption{Surfing transform analysis of the jetlet signals on July 3 2016. (a) Root mean square of the surfing signal a$_{s(t)}$ at different assumed propagation speeds; the peak is at v$_{peak}$=106 \kms. 
The nominal measurement error of the v$_{peak}$, as defined by the velocity step, is $\pm$2.2 \kms; a more realistic fitting error reflecting the flatness of the peak is likely around $\pm$10 \kms. (b) Waveform of the surfing signal a$_s$ at v = v$_{peak}$. (c) PSD of the waveform in panel (b) showing f$_{peak}$ $\approx$3.3 mHz.} 
\label{fig7}
\end{figure}

\begin{figure*}
\centering{
\includegraphics[width=16.4cm]{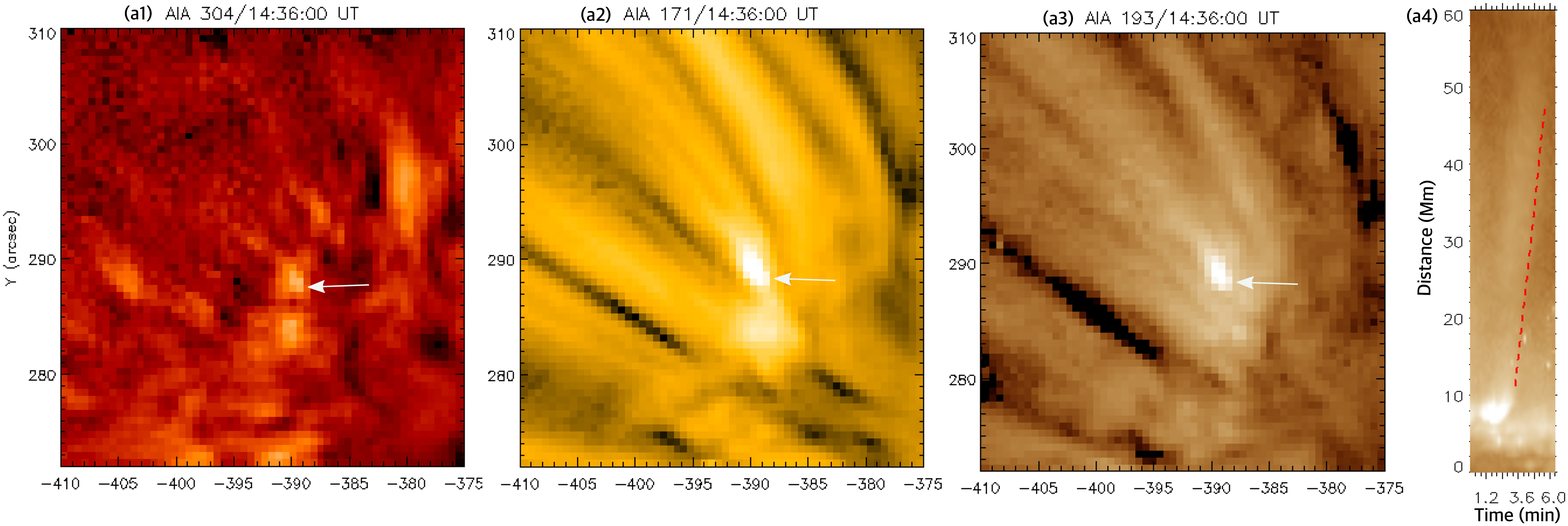}
\includegraphics[width=16.4cm]{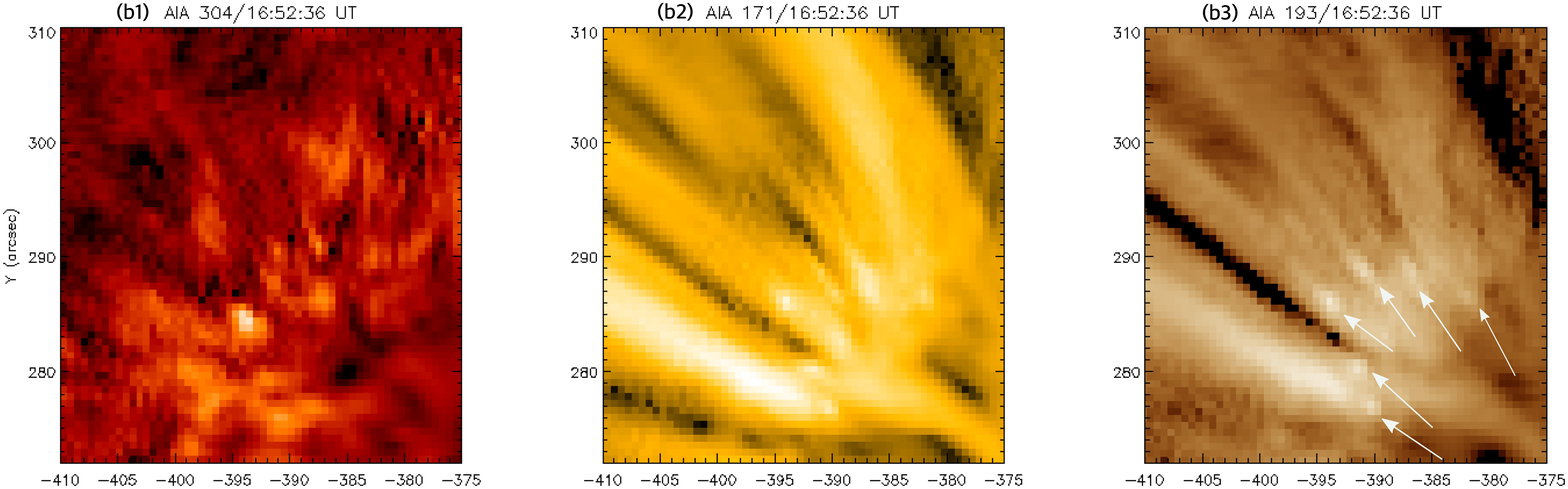}
\includegraphics[width=16.4cm]{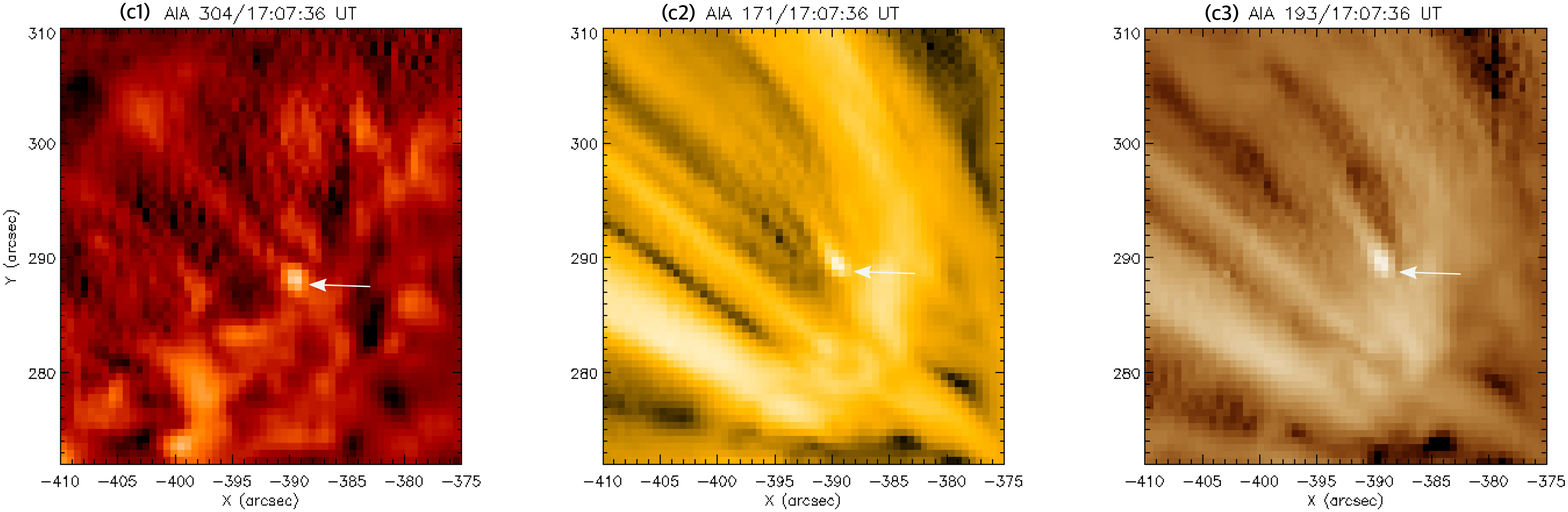}
}
\caption{Zoomed view of selected jetlets and associated brightenings at the base of P1 on July 3 2016. (a1-a3, b1-b3, c1-c3)  AIA 304, 171, and 193 \AA~ images showing multiple jetlets in the plume (marked by arrow). (a4) Zoomed view of a single-event time-distance intensity plot along the rectangular slits (red, marked in Figure~\ref{fig6}(a)) using 193 \AA~ intensity images. The start time is 14:33:36 UT. The red dashed line is a linear fit to the jetlet outflow (v$\approx$268$\pm$30 \kms). } 
\label{fig8}
\end{figure*}

\begin{figure*}
\centering{
\includegraphics[width=17cm]{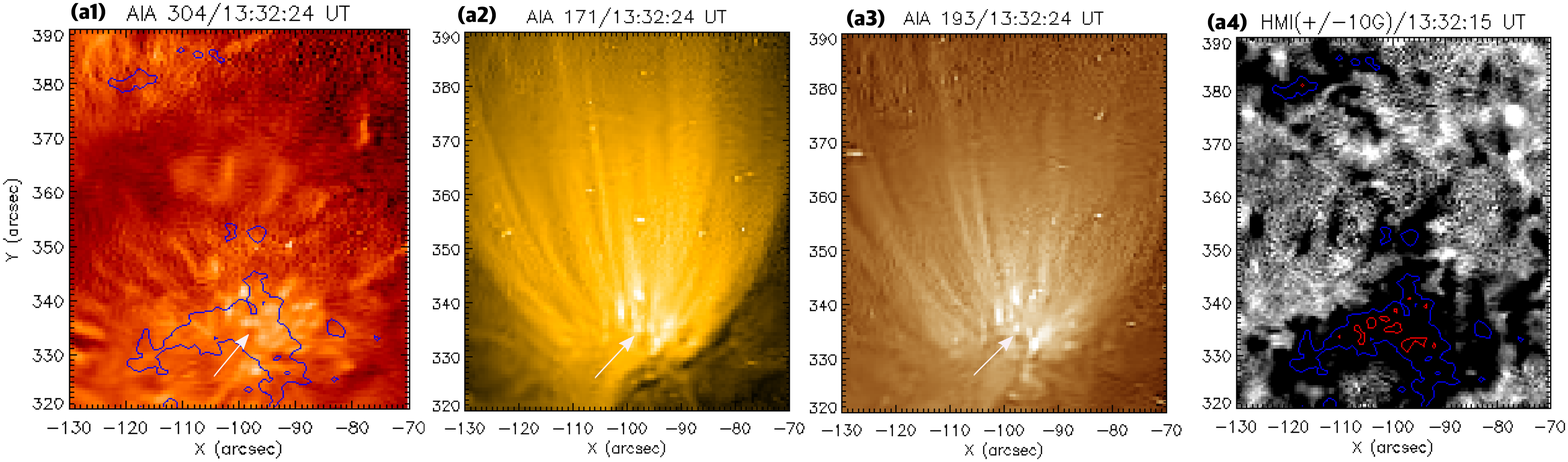}
\includegraphics[width=18cm]{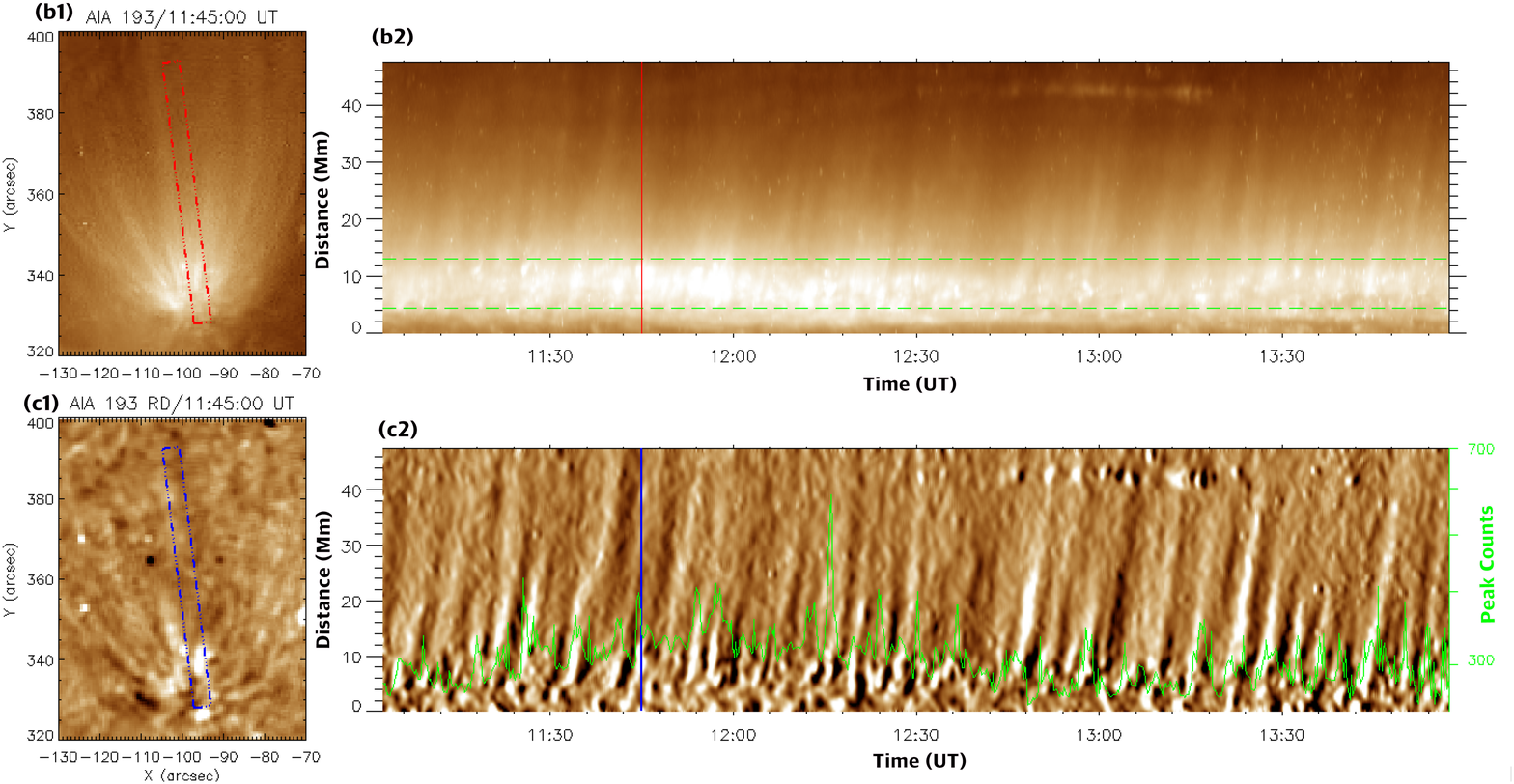}
}
\caption{Jetlets and associated brightenings at the base of plume P2 on Oct. 20 2015. (a1-a3)  AIA 304 (with -100 G magnetogram contours), 171, and 193 \AA~ images showing multiple jetlets in plume (one marked by an arrow). (a4) Cotemporal HMI magnetogram ($\pm$10 G). Blue and red contours indicate -100 G and -500 G photospheric fields. (b1,b2,c1,c2) Same as Figure \ref{fig6}. (An animation of this figure is available.)} 
\label{fig9}
\end{figure*}

\begin{figure*}
\centering{
\includegraphics[width=18cm]{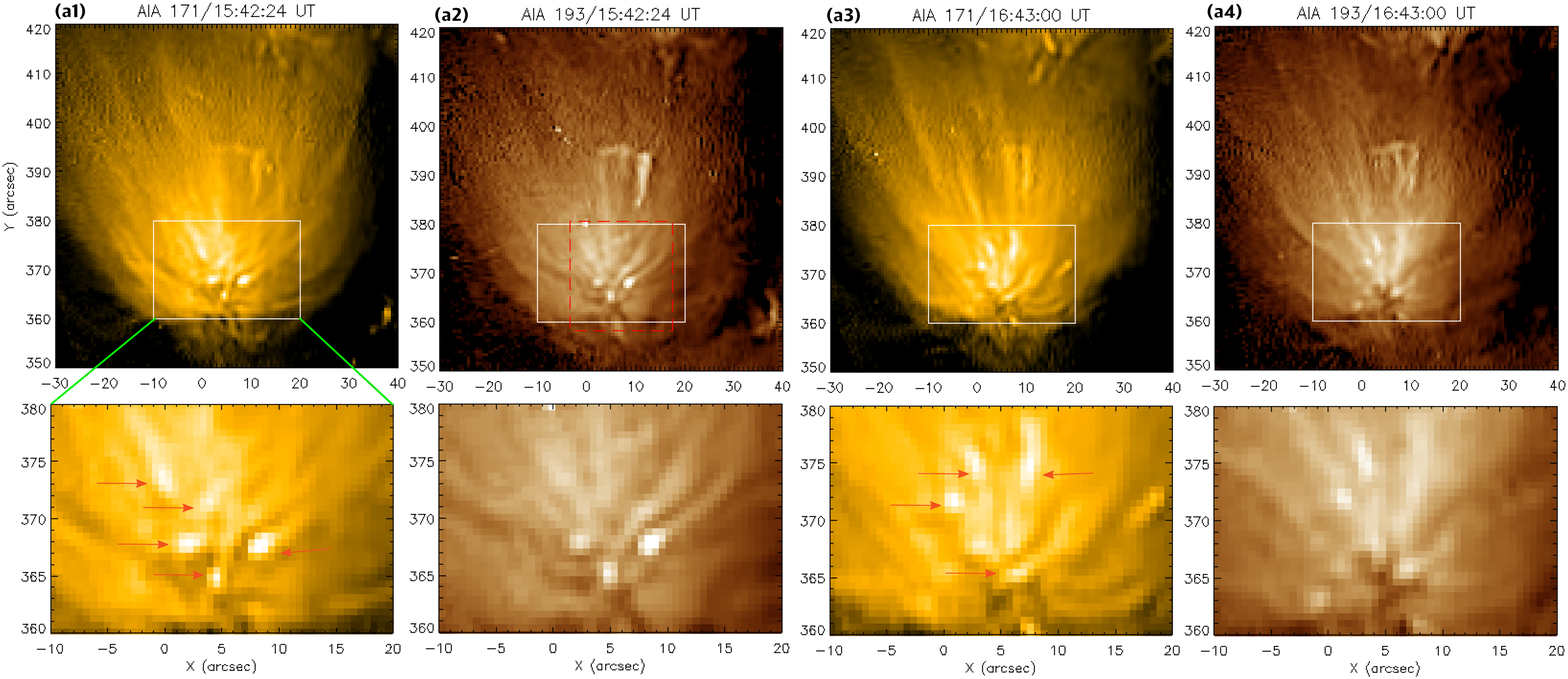}
\includegraphics[width=18cm]{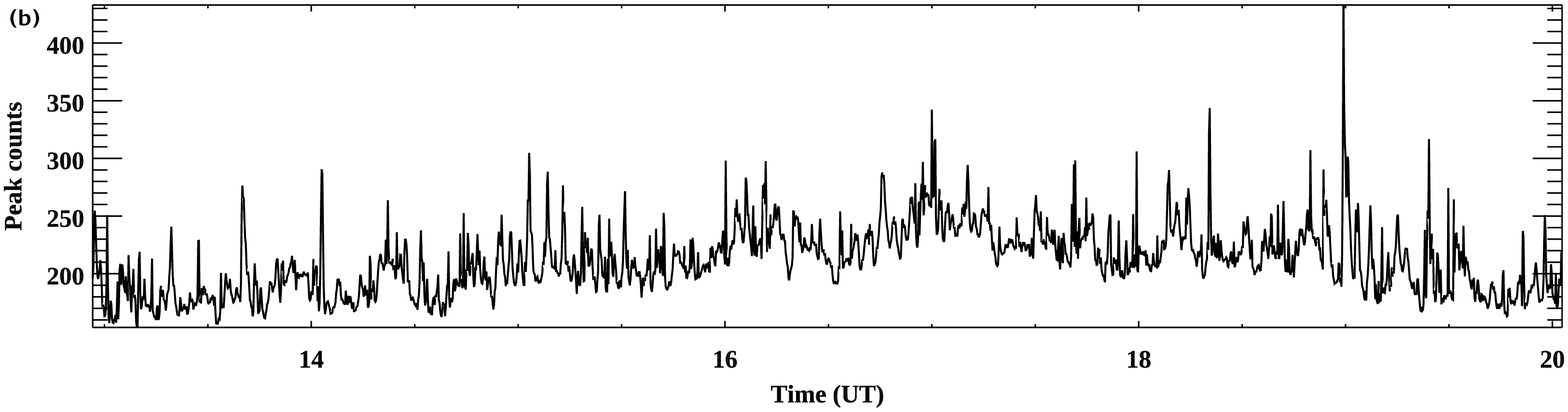}
}
\caption{Jetlets and associated brightenings at the base of plume P3 on March 19 2016. (a1-a4) 171 and 193 \AA~ images with the zoomed view of the footpoint region (rectangular box) showing multiple jetlets (marked by arrows).  (b) Peak 193 \AA counts extracted from the red box in (a2), showing quasiperiodic brightenings  ($\approx$3-5 min) at the plume footpoint.  (An animation of this figure is available.)} 
\label{fig10}
\end{figure*}

\begin{figure*}
\centering{
\includegraphics[width=16cm]{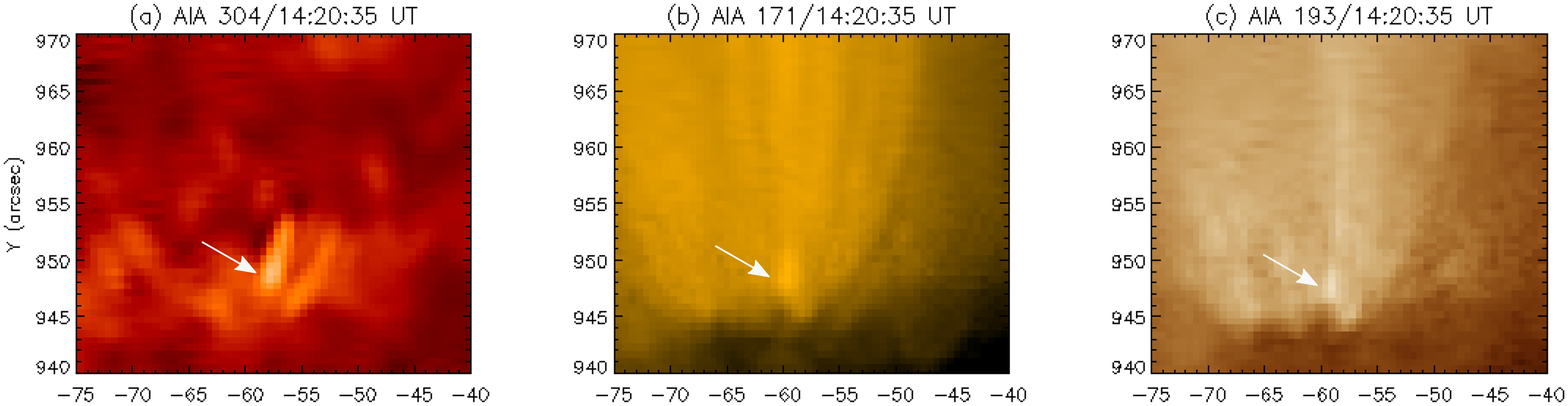}
\includegraphics[width=16cm]{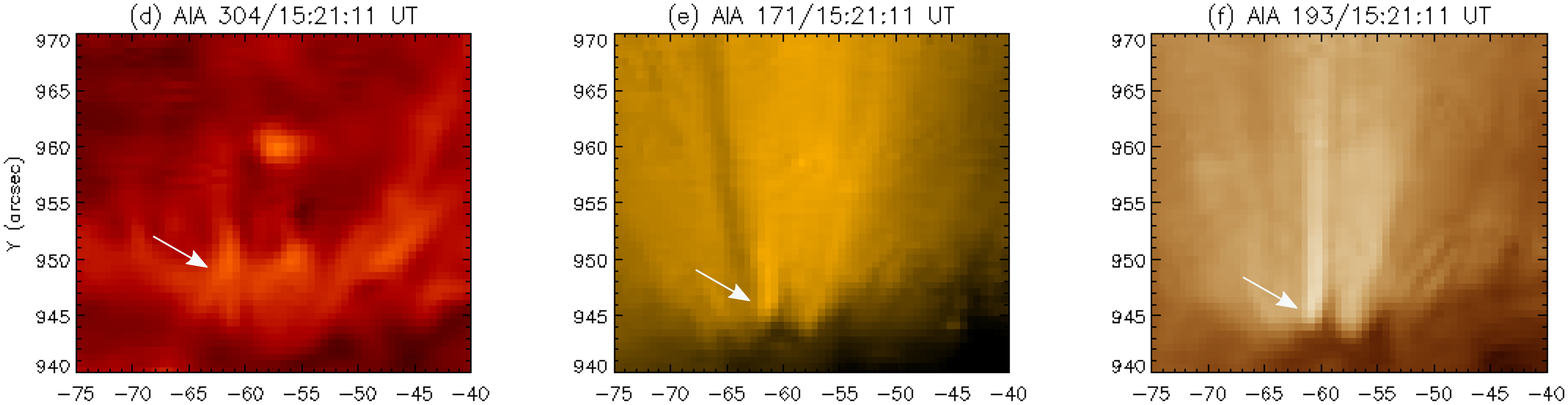}
\includegraphics[width=16cm]{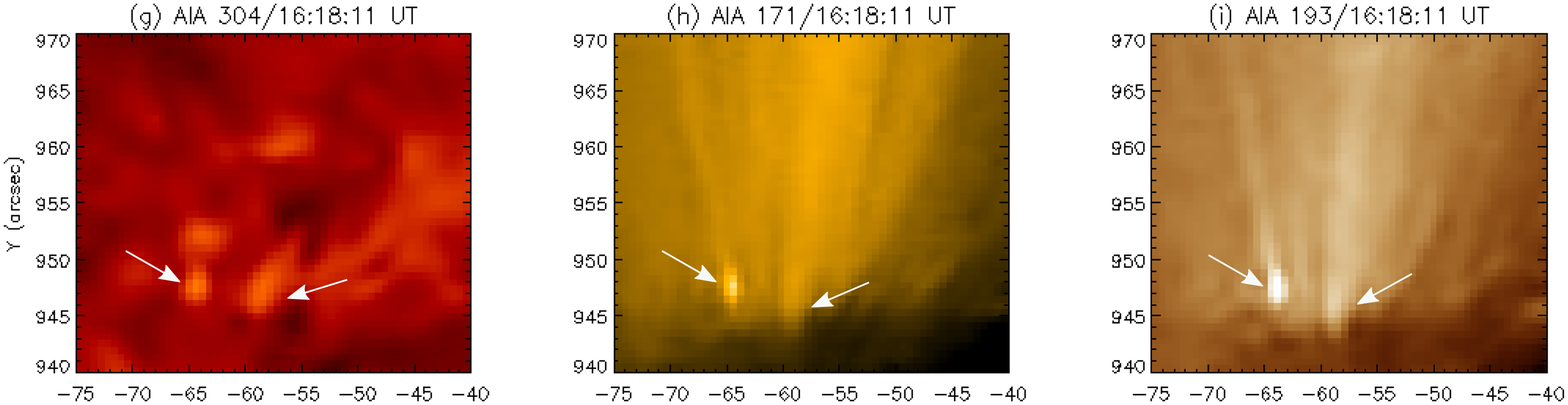}
\includegraphics[width=16cm]{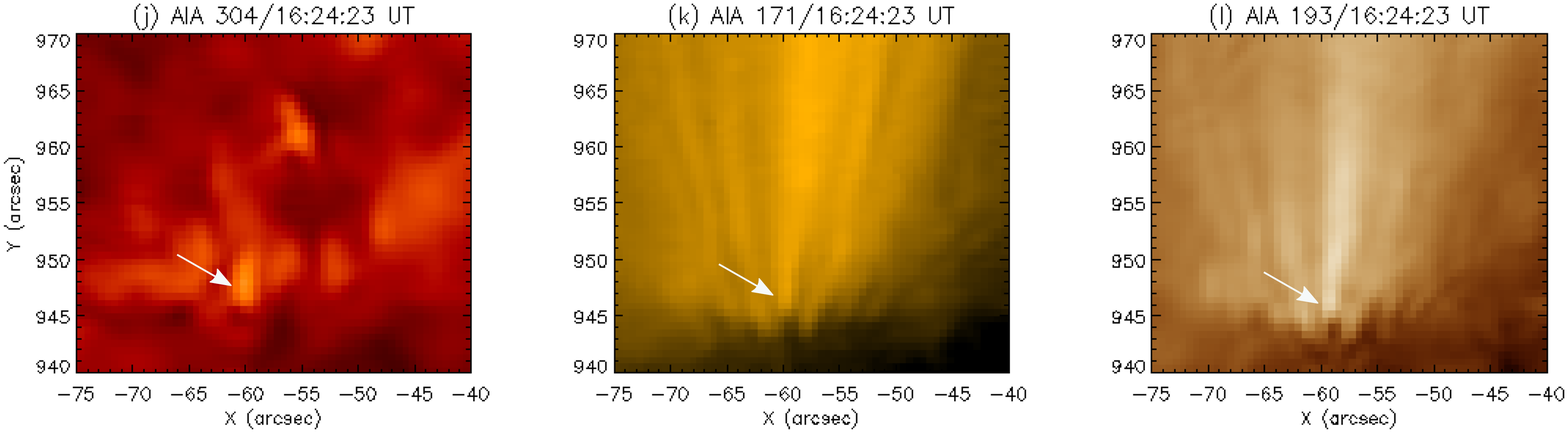}
}
\caption{Selected examples of jetlets and associated brightenings at the base of plume P4 on December 23 2019. (a-l) AIA 304, 171, and 193 \AA~ images showing multiple jetlets (marked by arrows) near the limb in the northern polar coronal hole. Panel (d) shows an inverse `Y' shaped structure. (An animation of this figure is available.)} 
\label{fig11}
\end{figure*}

\begin{figure*}
\centering{
\includegraphics[width=15cm]{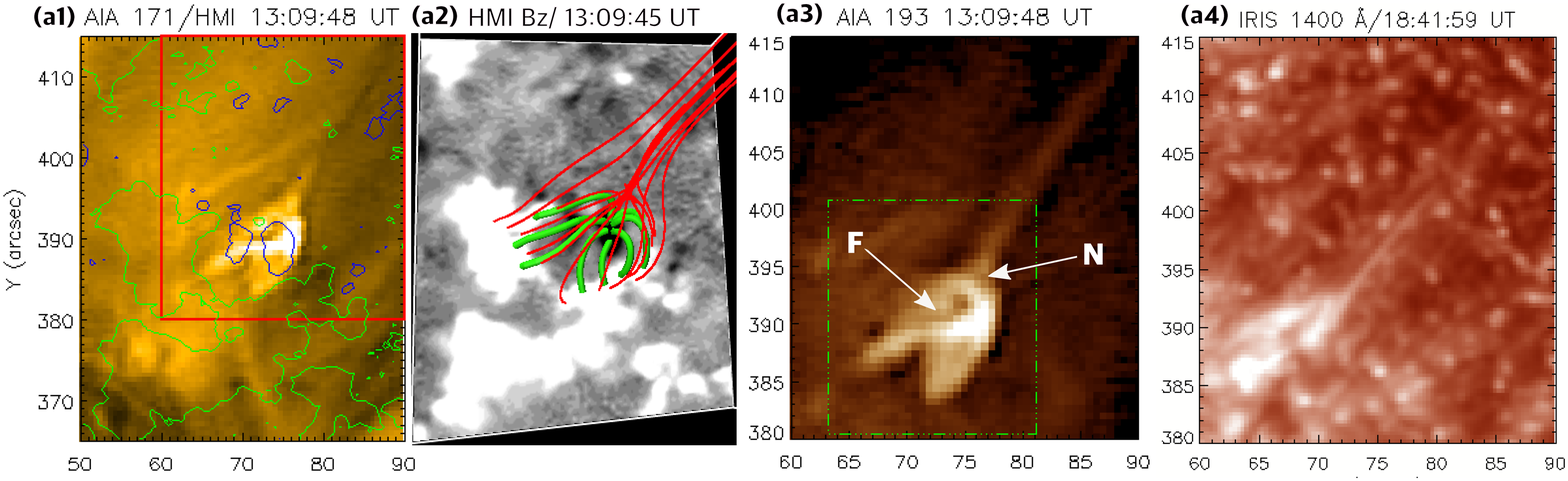}
\includegraphics[width=15cm]{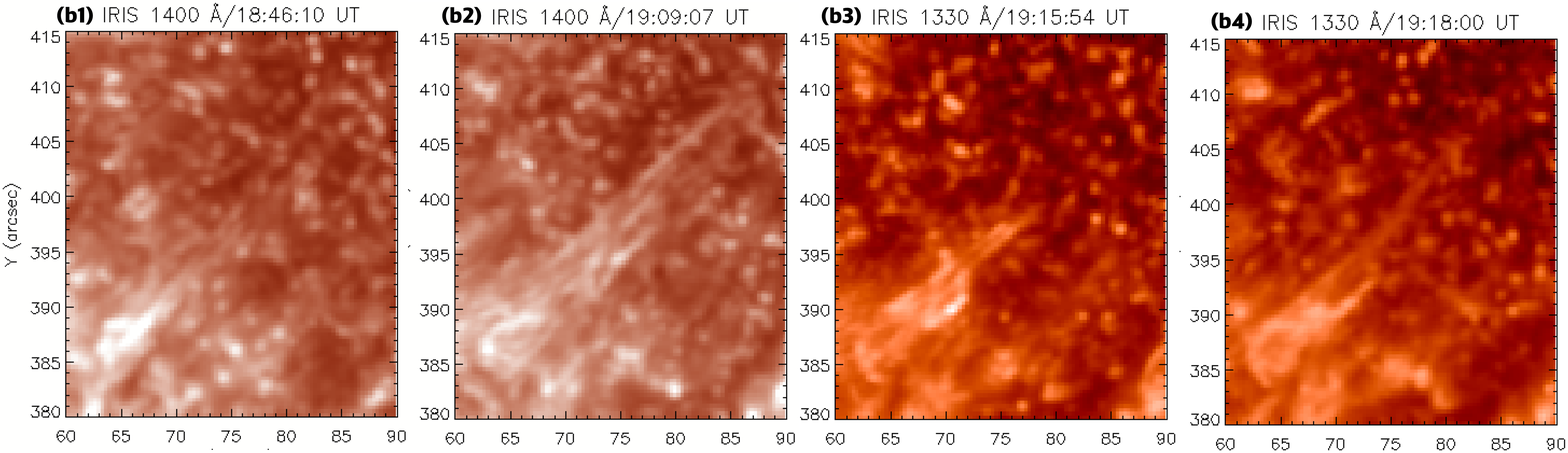}
\includegraphics[width=15cm]{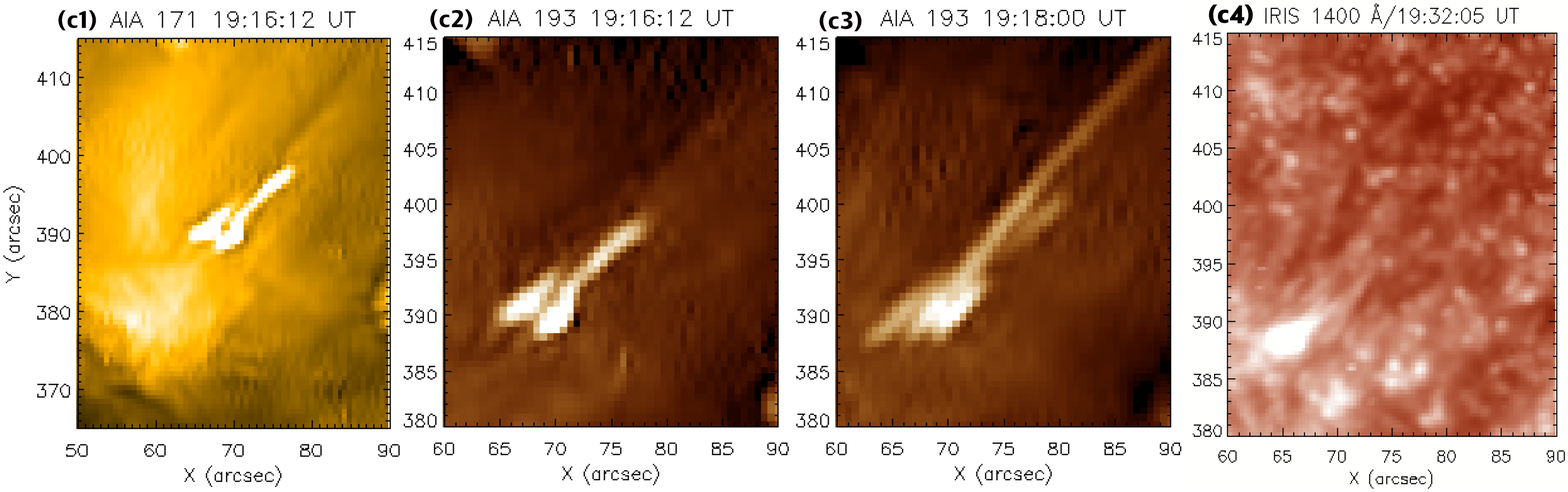}
\includegraphics[width=15cm]{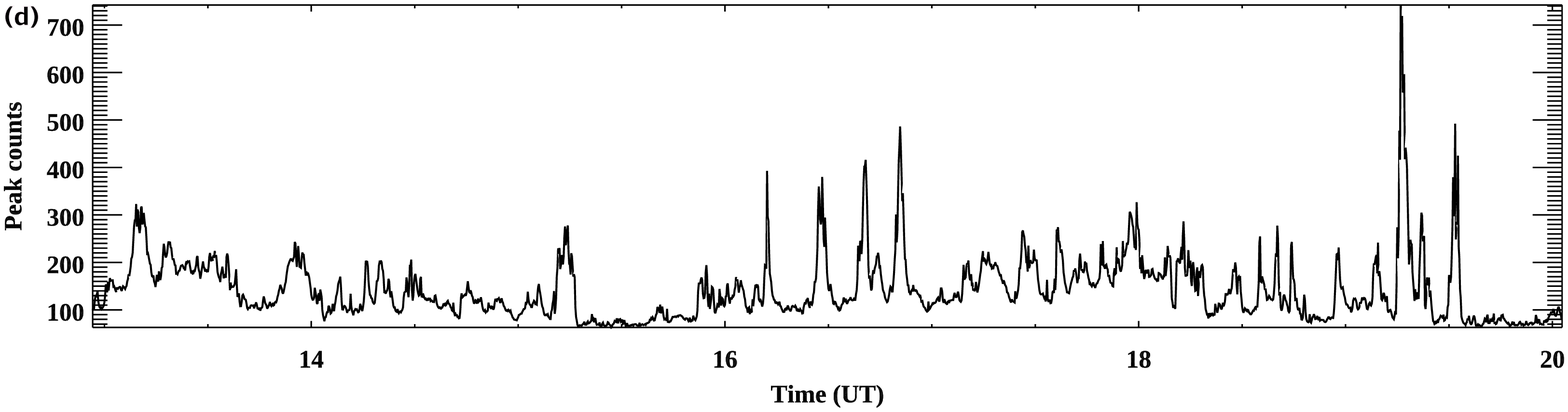}
}
\caption{AIA and IRIS observations of quasiperiodic jets from a small bright point (fan-spine topology) near the base of plume P3 in an equatorial CH on March 19 2016.  (a1) Selected AIA 171 image (overlaid by HMI magnetogram $\pm$10 G), (a2) Potential-field extrapolation (red/green indicate open/closed field lines.) (a3-c4) AIA 193 \AA~ and IRIS 1400 \AA\ slit-jaw images (zoomed view) showing the dynamic evolution of the jets. (d) Peak 193 \AA~ counts extracted from the green box shown in panel (a3), showing quasiperiodic brightenings (period $\approx$6-10 min) from the bright point. IRIS observed this plume (P3) during 18:19 UT to 21:48 UT. N=null-point, F=filament. (An animation of this figure is available.)} 
\label{fig12}
\end{figure*}

\begin{figure*}
\centering{
\includegraphics[width=12cm]{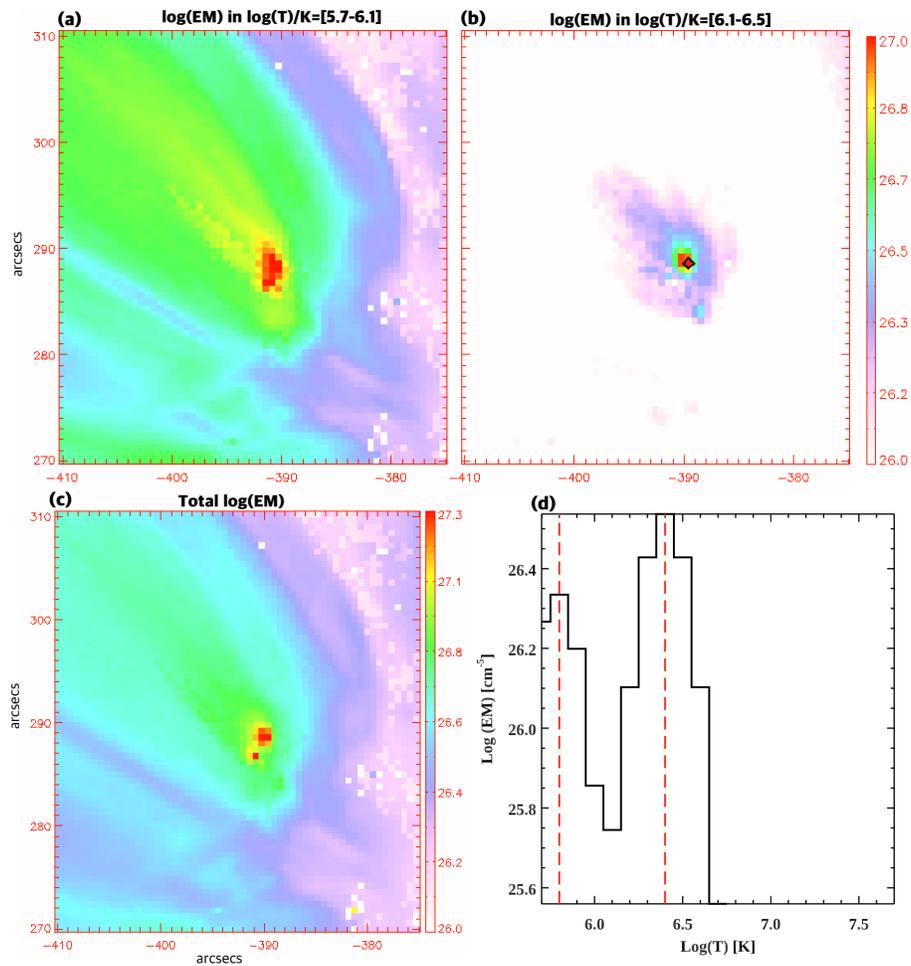}
}
\caption{(a-c) Emission measure (in cm$^{-5}$) maps obtained from the DEM analysis using images in six AIA channels at 14:36 UT (partly shown in Figure \ref{fig8}(a1-a3)). (d) DEM curve of the footpoint brightening (marked by a black diamond in panel (b)). The vertical dashed lines indicate DEM peaks at T=0.63 and 2.5 MK. } 
\label{fig13}
\end{figure*}

\begin{figure*}
\centering{
\includegraphics[width=13cm]{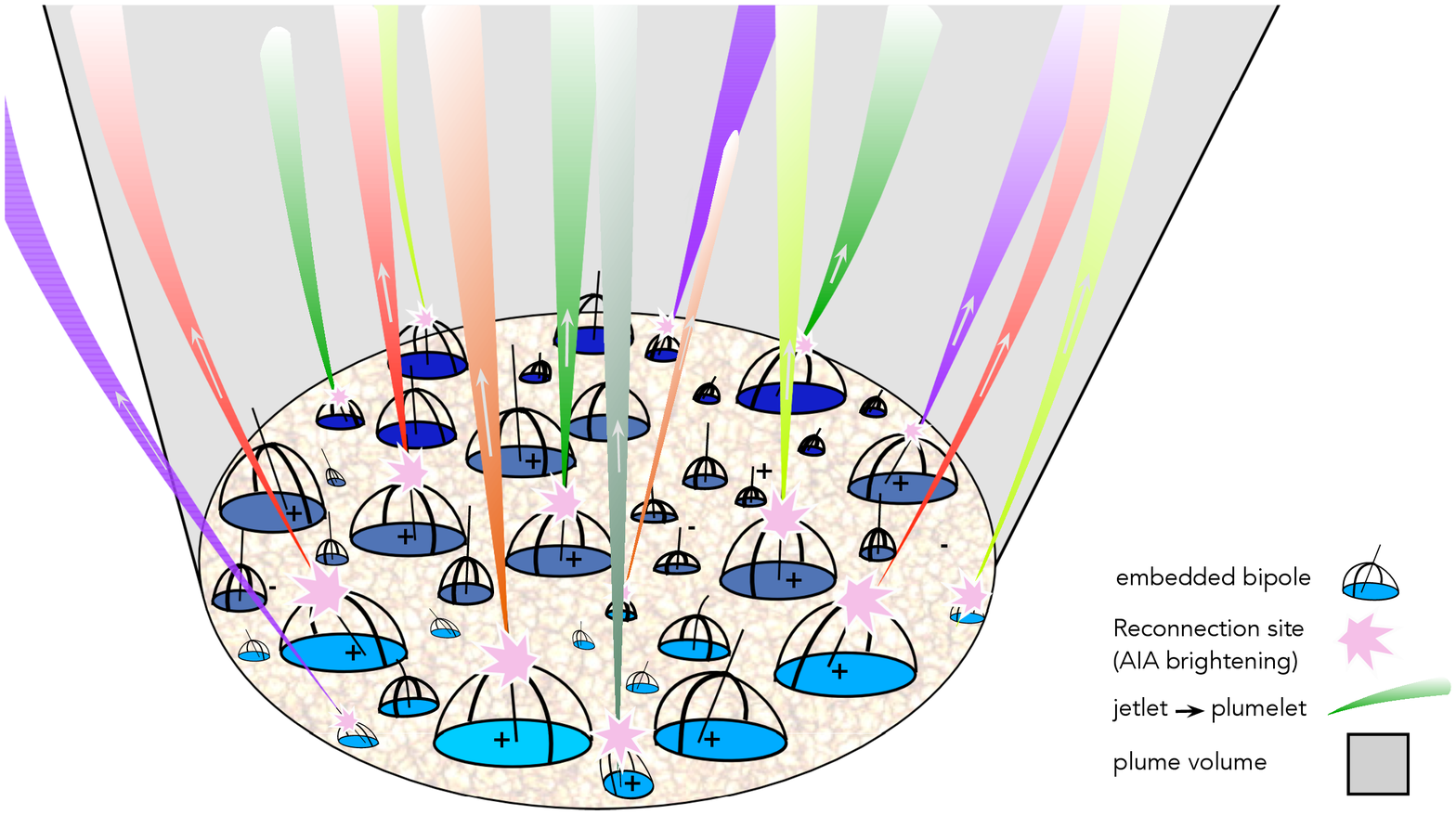}
}
\caption{Proposed scenario for the jetlet-plume relationship:  Tiny null-point (embedded-bipole) topologies at the base of a plume. Minority polarities in and around the majority polarities form fan-spine topologies, which release free magnetic energy through episodic reconnection at the nulls and possibly through and/or within the domes.  Quasiperiodic reconnection produces footpoint brightenings and outflowing jetlets (analogous to larger coronal-hole jets) that become plumelets, collectively comprising the plume, until the majority-polarity base disperses and the plume disappears.} 
\label{fig14}
\end{figure*}


\section{OBSERVATIONS}\label{obs}
We analyzed four plumes located inside near-equatorial and polar CHs. The coronal-hole environments for plumes P2, P3, and P4 are shown in Figure \ref{fig1}. All of the selected plumes are very bright and rooted in strong plages. Our analysis relied primarily on data from two {\it Solar Dynamics Observatory} (SDO) instruments. The Atmospheric Image Assembly (AIA; \citealt{lemen2012}) full-disk images of the Sun (field-of-view $\sim$1.3~R$_\odot$) have a spatial resolution of 1.5$\arcsec$ (0.6$\arcsec$~pixel$^{-1}$) and a cadence of 12~s, in the following channels: 304~\AA\ (\ion{He}{2}, at temperature $T\approx 0.05$~MK), 171~\AA\ (\ion{Fe}{9}, $T\approx 0.7$~MK), 193~\AA\ (\ion{Fe}{12}, \ion{Fe}{24}, $T\approx  1.2$~MK and $\approx 20$~MK), 211~\AA\ (\ion{Fe}{14}, $T\approx 2$~MK), AIA 94~\AA\ (\ion{Fe}{10}, \ion{Fe}{18}, $T\approx$1 MK, $T\approx$6.3 MK), and 131~\AA\ (\ion{Fe}{8}, \ion{Fe}{21}, \ion{Fe}{23}, i.e., 0.4, 10, 16 MK) images. We analyzed cotemporal SDO/Helioseismic and Magnetic Imager (HMI; \citealt{schou2012}) magnetograms at a 45-s cadence (noise level$\approx$5-10 G) to measure any photospheric magnetic-field changes during the jetlets. A 3D noise-gating technique \citep{deforest2017} was used to clean the SDO/AIA and HMI images. To determine the underlying magnetic structure of the investigated regions, based on the HMI magnetograms, we utilized a potential-field extrapolation code \citep{nakagawa1972} available in the GX simulator package of SSWIDL \citep{nita2015}. We also examined available Interface Region Imaging Spectrograph (IRIS; \citealt{bart2014}) slit-jaw images (125-s cadence, 0.33$\arcsec$~pixel$^{-1}$) of the jet source regions in the 1330 \AA\ (\ion{C}{2}, log T(K) = 3.7-7.0) and 1400 \AA\ (\ion{Si}{4}, log T (K) = 3.7-5.2) channels. IRIS observations in raster scan mode were available only for plume P3. IRIS slit-jaw images covered the entire plume (field of view=167$\arcsec \times 175\arcsec$, cadence=125 s) rooted in the curved plage/network region (Figure \ref{fig1}c,d). The IRIS field of view shifts from east to west during the observation.

We used the surfing-transform technique \citep{uritsky2013,uritsky2021} to determine the average phase speed and period of the jetlets. The average characteristic value of the phase speed was obtained by calculating the transform for a range of propagation speeds and determining which yields the strongest surfing signal.  The jetlet period was estimated by averaging the time intervals separating subsequent fronts in the image sequence, as well as from the position of the main peak in the slit-averaged Fourier spectrum. To reduce spectral noise, we averaged over three FFT-based power spectra obtained for three consecutive non-overlapping intervals covering the time of analysis. The intervals were preprocessed using the Hanning window function for eliminating edge effects and limiting spectral leakage.

The peak-count increase in most of the EUV footpoint brightenings (AIA 193/171 \AA) is about 1.5-2 times the background level; in some of the stronger events, the peak count is $\approx$3-4 times the background.  We performed a Morlet wavelet analysis \citep{torrence1998}  of the 193 \AA~ light curve to determine the period of oscillation (Figure \ref{fig4}). To obtain a detrended light curve (middle), the intensity profile (top) was first smoothed with a boxcar average of width 36 s, using the IDL SMOOTH routine, then a smoothed profile boxcar-averaged at 35 min (red) was subtracted. The wavelet power spectra (Fig. \ref{fig4}(c, d)) show that the periods of $\sim$3, 5-6, and 10 min are above the 99\% significance level.

\section{Results}\label{results}

 Plume P1 was located in the northern hemisphere (X,Y=-400$\arcsec$, 280$\arcsec$) inside an equatorial CH (Figure \ref{fig2}(a,b)), which extends to nearly one solar radius from east to west. The plume was rooted in a strong positive-polarity plage region (+400 G contours over AIA 171 \AA~ plume shown in Figure \ref{fig2}(c)) that matches the majority polarity. The HMI line-of-sight magnetogram reveals very weak minority polarity regions (-10 G) surrounding the dominant positive polarity (field strength contours +400, +200, +50 in Figure \ref{fig2}(d)). Note that we analyzed the higher portions of P1 in our recent study that discovered plumelets  \citep{uritsky2021}. 

An AIA 193 \AA~ animation (Figure \ref{fig3}) reveals multiple EUV brightenings rapidly appearing at random locations throughout the base of the plume, associated with outflows (jetlets) that transition smoothly into plumelets comprising the plume. We interpret these brightenings and jetlets as signatures of small-scale energy release. Similar brightenings with jetlets along the plume were also detected in AIA 171 \AA~ images, but they are best observed in the 193 \AA~ images (Figure \ref{fig3}(a,c)). An example of a tiny jet associated with brightening at a jetlet footpoint is marked by an arrow in Figure \ref{fig3}(a,c). To understand the temporal evolution of energy release at the plume base, we extracted peak counts within the red rectangular box in Figure \ref{fig3}a. The brightening associated with each energy-release episode forms a peak in the light curve (Figure \ref{fig3}(b,d)). Interestingly, the light curve shows quasiperiodic energy release during the entire period of observation (14:31:36 UT-17:27:24 UT).  A wavelet analysis of the smoothed and detrended 193 \AA~ light curve (Figure \ref{fig4}) identified peaks above the 99\% significance level at $\sim$3, 5-6, and 10 min.

Figure \ref{fig5}, and the accompanying 45-s cadence AIA movie overlaid by cotemporal HMI magnetogram contours, show the evolution of the photospheric magnetic field along with footpoint brightenings and jetlets. The minority polarities at the boundary of the dominant polarity are very weak (-10 G) (Figure \ref{fig5} top panels), so they are not seen within the $\pm$100 G contours (Figure \ref{fig5} bottom panels). The minority polarities at or outside this boundary move toward the dominant polarity and disappear after some interval (3-5 hours or less) depending on their sizes (see HMI magnetogram animation). Some brightenings at the plume base (marked by arrows) also originate roughly near the boundary, but most appear inside the +100 G field contours (Figure \ref{fig5}, bottom panels). Therefore we do not see consistent temporal correlations between the brightenings and disappearing minority polarities. Note, however, that the HMI resolution and sensitivity are not sufficient to resolve the minority polarities within the majority-polarity base of the plume. 

A time-distance (TD) plot of the 193 \AA~ intensity along a plumelet (red curved rectangular slit in Figure \ref{fig6}(a) was constructed to investigate the relationship between the footpoint brightenings and associated jetlets (Figure \ref{fig6}(b) and accompanying movie). We detected many tiny brightenings randomly distributed at the base of the plume, which precede the jetlets (see accompanying movie). The running-difference TD plot along the same rectangular slit reveals the jetlets more clearly (Figure \ref{fig6}(c,d)). The light curve (green curve in (d), from summed intensity between the green dashed lines in (b)) overplotted on the TD plot shows a clear correlation between individual footpoint energy-release episodes and jetlets. Our surfing-transform analysis  \citep{uritsky2009,keiling2012,uritsky2013} of the upward propagating jetlets yields a phase speed $v \sim$106 \kms and the frequency of the spectral peak (i.e., 3.26 mHz/5 min), which is very close to the typical frequency/period of the global p-mode oscillations (Figure \ref{fig7}). 
  
Figure \ref{fig8} depicts clear examples of P1 jetlets detected in the AIA 304, 171, and 193 \AA~ channels. The top panels (Figure \ref{fig8}(a1-a3)) reveal one of the strongest jetlets at 14:36 UT. The corresponding footpoint brightening is also seen in the AIA hot channels (131/94 \AA). The 193 \AA~ TD plot shows a zoomed view of this footpoint brightening and associated jetlet; the dashed line represents a projected speed of $v \approx$268$\pm$30 \kms (Figure \ref{fig8}(a4)). Jetlets occur at several different footpoints within the plume base (16:52:36 UT, Figure \ref{fig8}(b1-b3)).  

Similar quasiperiodic jetlets were observed in plume P2 (X=-100$\arcsec$,Y=350$\arcsec$),  located in an equatorial CH on October 20 2015. A group of footpoint brightenings (marked by an arrow) associated with jetlets is clearly seen in AIA 171 and 193 \AA~ (Figure \ref{fig9}(a1-a3) and accompanying movie). HMI magnetograms show weak minority polarities (-10 G) at the boundary of the majority-polarity plage (Figure \ref{fig9}(a4)). With the above caveat about insufficient HMI sensitivity and spatial resolution, neither flux emergence nor submergence/cancellation was observed in correlation with jetlets (see accompanying movie).  Other aspects are consistent with plume P1: jetlets appear near the boundary and within the -100 G contour; the TD plot along the red rectangular box (Figure \ref{fig9}(b1,b2)) reveals quasiperiodic jetlets rising from footpoint brightenings that repeat every 3-5 min (Figure \ref{fig9}(c2)); the peak emission intensity correlates well with the associated jetlets (Figure \ref{fig9}(c2)). Our surfing-transform analysis (not shown) provides an average projected speed for the P2 jetlets of $v \approx$65 \kms, with a period of $\approx$5 min.   

Plume P3, located near the disk center in an equatorial CH on March 19 2016 (X=5$\arcsec$, Y=370$\arcsec$) contains similar multiple jetlets associated with quasiperiodic brightenings at the base of the plume for $\sim$7 hours. The zoomed views of the footpoint regions exhibit multiple brightenings/jetlets (width$\approx$2-3$\arcsec$) at 15:42:24 UT and 16:43 UT (Figure \ref{fig10}(a1-a4) and accompanying movie). The peak counts extracted from the footpoint region show significant fluctuations associated with episodic energy-release in the jetlets with a $\approx$3-5 min period (Figure \ref{fig10}(b)). 

In general, it is difficult to see the chromospheric counterparts of jetlets in on-disk plumes due to projection effects, the small jetlet size, and the limited AIA spatial resolution. Figure \ref{fig11} and accompanying movie show chromospheric counterparts of selected brightenings and associated narrow jets propagating along plume P4 in the northern CH (near the limb). The jetlet seen in the 304 \AA\ channel only extends a short distance into the corona, whereas the hotter components, best seen in the AIA 193 \AA\ channel, clearly propagate much higher to form a plumelet. The characteristics of the footpoint brightenings and jetlets in plume P4 match those of their counterparts in plumes P1-P3, as reported above (Table 1). Table 1 summarizes the results for all plumes (P1-P4) using the same wavelet analysis for footpoint energy release (EUV brightenings) and surfing transform analysis technique for the jetlets extending upward from the EUV brightenings along the plume.     

\section{DISCUSSION}\label{discussion}

\subsection{Magnetic topology and generation mechanism of jetlets}
The average width of brightenings within the plume base is around 2-3$\arcsec$, which is at the limit of the SDO/AIA spatial resolution.  Therefore these EUV observations could not resolve the plasma structure that traces the magnetic topology beneath most of the jetlets. In addition, most HMI observations are unable to detect or resolve the minority polarities at the base of the analyzed plumes. However, some of the larger brightenings (dome width$\approx$4-6$\arcsec$) in these plumes are resolved in AIA and high-resolution IRIS observations. We analyzed the magnetic topology of a big jetlet that is about two times larger than a typical jetlet but smaller than typical jets \citep[e.g.,\ ][]{kumar2019a}.  For plume P3 on March 19 2016, weak minority polarity located near the edge of the majority-polarity region is visible in HMI magnetograms (Figure \ref{fig12}(a1,a2)). The potential-field extrapolation reveals a fan-spine topology with a 3D null (Figure \ref{fig12}(a2)). The zoomed view of this region in the AIA 193 \AA~ channel at 13:09:48 UT shows a dark mini-filament seen in absorption ($\approx$1-2$\arcsec$ wide), associated brightening below the null, and collimated outflow from the null (Figure \ref{fig12}(a3)). Interestingly, the AIA 171/193 \AA~ animation (Figure \ref{fig12}) shows quasiperiodic brightenings (period$\approx$6-10 min) and  jetlets for about 6 hours until the disappearance of the minority polarity and shrinking EUV bright structure (Figure \ref{fig12}(d)). These results are consistent with the lifecycle of the source regions (coronal bright points) in our statistical study of jets in an equatorial coronal hole  \citep{kumar2019a}. 

IRIS slit-jaw images of the transition region (TR) in 1400 \AA/Si IV at 18:41:59 UT show another jet from the same fan-spine configuration, with a quasicircular brightening at the footpoint (Figure \ref{fig12}(a4)). \citet{panesar2018} also reported IRIS observations of a jet (19:15-19:20 UT) from a similar source region and found some decrease in minority-polarity flux in HMI.  AIA 171 and 193 \AA~ images reveal dome brightening and associated blob-like structure (i.e., a plasmoid) in the jetlet (Figure \ref{fig12}(c1,c2)). An extended jet along the outer spine appears 1-2 min later in both IRIS 1330 \AA~ and AIA images (Figure \ref{fig12}(c3,b4)). 

IRIS and AIA images clearly show brightenings at the base of the dome during the jetlet initiation phase, analogous to the circular ribbons seen in bigger jets in active regions and CHs  \citep{masson2009,wang2012,lee2020,kumar2018,kumar2019a}. This is a clear signature of interchange reconnection at the stressed null between the closed flux beneath the dome and the adjacent open flux, a key element of the breakout jet model discussed above and a common feature of jets \citep{kumar2018,kumar2019a,kumar2019b}.  The broader brightenings beneath the dome, only resolvable for the largest jetlets, are similarly analogous to the flare arcades seen in observations and models of jets and CMEs. Therefore we speculate that tiny null-point topologies (dome width$\approx$2-3$\arcsec$) scattered throughout the plume base undergo repeated episodes of reconnection, releasing jetlets along the spine and heating the plasma inside the dome (see Figure \ref{fig12}). From the base/dome width, we roughly estimate the height of the null to be around $\approx$2-3$\arcsec$   \citep{kumar2019b, kumar2021}. Therefore we suggest that most of these jetlet-driving reconnections occur in the upper chromosphere/TR ($\approx$1500-2200 km), as confirmed by the high-resolution IRIS observations reported here for P3 (Figure \ref{fig12}).

The quasiperiodic nature of these episodes merits further consideration.  The peak of the intensity power spectrum (Figure \ref{fig4}) is remarkably close to the dominant p-mode period, and comparable periodicities have been detected in other explosive/eruptive phenomena  \citep{ning2004,doyle2006,sych2009,gupta2015,kumar2015,kumar2016}.   Photospheric p-modes can penetrate to the starting height of the jetlets, i.e. chromosphere/TR and higher, along oblique field lines \citep{depontieu2004,depontieu2005,heggland2011,khomenko2019}. Transverse motions with comparable periods also have been detected in the northern polar coronal hole and large coronal loops \citep{tomczyk2009,morton2015}. It is unclear, however, whether the jetlet reconnection itself is driven periodically by the p-mode waves rising from the solar interior, or whether the observed oscillations simply modulate the jetlet flows.

In the first scenario, the jetlets reported here result from periodic reconnection, which generates periodic plasma flows along the plume.  In general, null-point oscillations can trigger oscillatory reconnection  \citep{thurgood2019}. Specifically, the global p-mode waves in the chromosphere/TR could distort or stress the null into a current sheet and cause repetitive reconnection episodes \citep[e.g., ][]{chen2006}. In the second scenario, the jetlets are generated by intermittent localized episodes of reconnection, but the surface motions caused by the p-modes drive the jet sources up and down, generating waves that propagate into the corona as the jetlets transform into plumelets. We have already reported that plumelets oscillate with the same range of periods as the jetlets \citep{uritsky2021}. 

\subsection{The controversy of outflows vs waves in plumes}
PDs in plumes are widely interpreted as either plasma outflows or slow-mode waves \citep{wang2016}. The observations reported here resolve this controversy in a conciliatory way: both strong flows and waves coexist in the jetlets and the overlying plumelets. Reconnection at the plumelet footpoints drives flows that are observable as jetlets, as well as heating the plasma in the tiny embedded-bipole sources. Reconnection alone frequently generates Alfv\'enic fronts, as demonstrated by our simulations of larger-scale coronal jets \citep{karpen2017}. The dense material in the jets, however, travels much more slowly, closer to the coronal sound speed. Between the jet and the Alfv\'enic front lies an inhomogeneous wake of shear and compressible turbulence \citep{uritsky2017}. Therefore we conclude that the jets are outflows, not slow-mode waves, but the underlying reconnection and flows can generate various waves along the newly reconnected field lines.  MHD simulations have shown that quasiperiodic upflows at the base of coronal loops can generate slow-mode waves \citep{ofman2012,wang2013} and shocks \citep{petralia2014}. These calculations do not include magnetic reconnection, however, which introduces other types of waves onto the reconfigured field lines.  Our observations and the abovementioned simulations confirm that slow-mode and Alfvénic waves could coexist with reconnection-generated outflows, but do not definitively establish whether the p-mode waves directly drive periodic reconnection by perturbing the overlying null point or modulate jetlet outflows from impulsive reconnection by vertical oscillations of the underlying magnetic structure. Further observational and computational work is needed to resolve this question.

\subsection{Jetlet periodicity}
We have confirmed the 5-min period in a single plumelet with two different methods: wavelet (energy release at the base) and Uritsky’s surfing-analysis technique (upward propagating jetlets along the plume, see Figure \ref{fig7}). Furthermore, in the AIA time-distance plot, we can see clearly the intensity variations every $\approx$5 min associated with jetlets, even without performing any detrending or wavelet analysis. \citet{uritsky2021} already analyzed multiple intervals during the P1 observing sequence and found a similar 5 min (3.3 mHz) period in several plumelets, after the early stage mentioned above. Our investigation does not rule out the coexistence of longer periods at higher heights in plumes, however. In fact, plume P1 exhibited longer periods at 12 min and 24 min during the early stage of plume formation, when the plume was faint  \citep{uritsky2021}.

The plumes that we studied were rooted in very strong plages, and were young when their jetlet activity reached its peak. Longer periods ($\approx$10-30 min) have been observed with AIA in polar CH plumes that were very faint and possibly rooted in weaker magnetic flux \citep{krishna2011,Krishna2014,jiao2015,yuan2016,cho2021}. In addition, transient events at the base of the plumes were not reported in those earlier studies. Based on these results, we speculate that strong, bright plumes manifest more jetlet activity than fainter plumes rooted in weaker magnetic field, but a more thorough investigation is needed to test this conjecture.
   
\subsection{Jetlet thermal and kinetic energy}
We utilized an open-source differential emission measure (DEM) code \citep{cheung2015} to determine the densities and temperatures for one of the strongest jetlets observed in P1. The code uses six AIA-channel images (94, 131, 171, 193, 211, 335 \AA) as input to calculate the DEM. The EM maps (in units of 10$^{26}$cm$^{-5}$) at different temperatures are shown for this strong jetlet at 14:36:00 UT (Figure \ref{fig13}(a-c)). 

To determine the density of the jetlet and associated brightening near the base of plume P1, we first estimated the total EM by integrating the DEM distribution over the entire Gaussian temperature range within the jetlet ($\approx$7.6$\times$10$^{26}$ cm$^{-5}$) and brightening ($\approx$1.2$\times$10$^{27}$ cm$^{-5}$) (Figure \ref{fig13}c). The DEM analysis of the brightening shows double peaks at temperatures $T \approx$0.63 and 2.5 MK (Figure \ref{fig13}d) . Assuming that the depth of the structure along the line of sight is approximately equal to its width $w$, then the densities of the jetlet and associated brightening are n =$\sqrt{\frac{EM}{w}}$ = 2.3$\times$10$^9$cm$^{-3}$ and 2.8$\times$10$^9$cm$^{-3}$, respectively (assuming filling factor$\approx$1). $w \approx$2$\arcsec$, so the estimated thermal energy (E=3Nk$_B$T) 9.6$\times$10$^{24}$ ergs for 2.5 MK, where the total number of electrons (N) in volume V ($\approx$w$^3$) is nV. 

Assuming a cylindrical jetlet of radius $\approx$1$\arcsec$ and a length of $\approx$5$\arcsec$ (lower limit) from our observation, the jetlet mass would be $m \approx$2.5$\times$10$^{10}$ g. Therefore, the estimated kinetic energy KE=$\frac{1}{2}$mv$^{2}$ $\approx$ 1.3 $\times$10$^{24}$ ergs for an average speed of 100 \kms. The estimated jetlet thermal energy is an order of magnitude higher than its kinetic energy; the latter is equivalent to the estimated energy of a nanoflare \citep{parker1988}. Because this estimate is for one of the strongest jetlets observed, however, it should be considered an upper limit; most jetlets should have smaller thermal and kinetic energies.

\subsection{Mass loss rate from jetlets}
We estimate the mass flux from the jetlets as m=4$\pi$r$^2$$\rho$vf$_{CH}$f$_P$f$_J$, where r=1.0 R$_{\odot}$, $\rho$ is the mass density (1.1$\times$n$_{e}$m$_{p}$, n$_{e}$=electron number density, m$_{p}$=proton mass), the jetlet speed v$\approx$100 \kms, f$_{CH}$=0.1 (coronal-hole area relative to the solar surface$\approx$10$\%$), f$_{P}$$\approx$0.1 (plumes covering roughly 10$\%$ of the CH area), and f$_J$ is the filling factor for jetlets/plumelets within plumes. 

We use the results of our plumelet analysis \citep{uritsky2021} to estimate f$_J$. Suppose a plume averages N$_J$ jetlets/plumelets at any time, the average plumelet width is w, and the plume width is W.  Then the volume filling factor is f$_J$ = N$_J$ (w/W)$^2$. For plume P1,  N$_J$ = 10 and w = 10$\arcsec$, and the plume width W is 150$\arcsec$ (Figure 2), yielding f$_J$ = .05. 

The estimated mass flux is $\approx$1.2$\times$10$^{12}$ g s$^{-1}$, which is lower than the mass flux from network jets derived from IRIS observations  \citep{tian2014}, but higher than the rough estimates of mass flux from a polar plume and open structures at the periphery of an active region \citep{sakao2007,cho2020}. Therefore the jetlet mass flux is roughly comparable to the mass-loss rate in the wind at solar minimum, but a smaller fraction ($\approx$ 60$\%$) of the solar-maximum wind \citep{wang1998,wang2020}. Note that this is an upper limit, however, because the amount of mass that falls back to the surface during a jetlet is unknown.  Judging from observations of jets \citep{kumar2019a} and Type II spicules/network jets \citep{tian2014,samanta2015}, most of the cooler, denser plasma is likely to return. But we expect the warm plasma ($\approx$1 MK) in jetlets to escape. More multi-wavelength imaging of jetlets with high resolution and cadence, and spectroscopic observations of jetlets on the disk, are needed to determine the fraction of escaping plasma. On the basis of these crude estimates, jetlets might contribute substantially to the solar wind.

\subsection{A possible connection to magnetic switchbacks}
The quasiperiodic variations of radial magnetic field and velocity components in magnetic switchbacks \citep{kasper2019,bale2019,horbury2020} recently detected by Parker Solar Probe (PSP; \citet{fox2016}) have sparked great interest in the heliophysics community, and multiple explanations have been offered. Interchange reconnection has already been proposed as a likely candidate for the magnetic switchbacks \citep{zank2020,fisk2020,sterling2020a,sterling2020b}. The estimated scale-size of PSP magnetic switchback clusters is equivalent to the size of a supergranule \citep{bale2021,Fargette2021}, leading to the suggestion that switchbacks originate from magnetic funnels associated with the network magnetic field via interchange reconnection events \citep{bale2021}. 

The present study leads us to an alternative hypothesis. Coronal holes are filled with numerous bright points of different sizes, which are well-known sources of repeated coronal jets produced by flare and interchange (breakout) reconnection \citep{kumar2019a}. A bright point forms when an embedded bipole emerges in a background majority-polarity region, yielding a fan-spine topology with a 3D null at a height comparable to the dome width \citep{wyper2017,wyper2018,kumar2019a}. In fact, the tiny bright points associated with brightenings at the bases of plumes are similar in size to the widths of supergranule boundaries, where mixed-polarity magnetic fields collect and interact. 

If PSP is well-connected magnetically to these small bright points at plume bases, which undergo repetitive reconnection episodes, then jetlets are likely candidates for the solar origins of the magnetic switchbacks detected in the solar wind. 

Microstreams \citep{neugebauer1995} are fluctuations in the solar wind speed/density associated with folds in the magnetic field (the original phenomenon denoted switchbacks), emanating from polar coronal holes \citep{thieme1990} where the proton velocity was $>$20 \kms above or below the running-average speed for an average duration of 0.4 days \citep{neugebauer2021}. These structures were  detected by Helios at 0.3 AU and Ulysses (between $\approx$1 and 3 AU) in the fast solar wind from polar coronal holes \citep{thieme1990,neugebauer1995}. The lifetime of microstreams can be roughly equivalent to the lifetime of repetitive jets in coronal bright points in and around plumes. It has been suggested that they are likely generated by jets \citep{neugebauer2012,neugebauer2021}, so quasiperiodic interchange reconnection associated with jetlets from coronal bright points also could explain microstreams.

In future, to establish whether a direct connection exists between switchbacks and interchange reconnection in the chromosphere/TR beneath plumes, we need more observations that connect the PSP in-situ observations of switchbacks to their source regions at the Sun. In addition, MHD simulations following breakout jetlets from the Sun to PSP locations would test this conjecture by determining whether an interchange/breakout reconnection-generated kink in the open field lines can propagate to PSP distances.

 \section{CONCLUSIONS}\label{conclusions}
We discovered numerous repetitive brightenings with periods of 3-5 minutes, randomly distributed in space at the bases of several plumes in equatorial and polar coronal holes. These small-scale energy release produce plasma outflows denoted ``jetlets", which transition smoothly into bright ``plumelets" that comprise the overlying plumes.  The coronal manifestations of this process are best seen in the AIA 193 \AA~ channel (T$\sim$1.2 MK), while 304 \AA~ channel images show the chromospheric/transition-region counterparts of the footpoint brightenings and jetlets. Both theory and observations show that reconnection at a 3D null in the upper chromosphere or transition region can precipitate energetic electrons to the chromosphere and eject hot plasma outward. For example, a 3D radiative-MHD numerical simulation has shown episodes of plasma jets and heating over 1 MK via null-point reconnection in the TR induced by upward propagating waves \citep{Heggland2009}.
Heating beyond 1 MK is also observed in numerous  larger jets and flares during explosive breakout and flare reconnection as well as the associated evaporation. We have demonstrated that plumelets originate in jetlets, by tracing localized flows from the base into the plume above.  According to our observations and simulations, slow bulk upflows can coexist with Alfv\'enic fronts in plumes, although they separate rapidly with time. Therefore we conclude that the slower, denser jetlets are indeed outflows, while PDs and other wave phenomena could be driven ahead of the jetlets. 

HMI magnetograms reveal that small, weak (10-20 G), minority-polarity magnetic features converge on the boundary of the stronger (100-200 G), majority-polarity region in which the plumes are rooted. These parasitic polarities then disappear, presumably via submergence/cancellation with the strong majority-polarity flux. However, few if any of the observed jetlets and associated brightenings are temporally and spatially linked to these interactions. Inside the majority-polarity patch, HMI does not show any evidence of minority-polarity intrusions. However, high-resolution NIRIS magnetograms from Goode Solar telescope (GST) \citep{samanta2019,abramenko2020} demonstrate that suintrusions generally exist, but are below the threshold for HMI detection. In fact, most of the jetlets and associated brightenings occur within the strong plage region, rather than at the boundaries. There is no evidence in the HMI data for repetitive emergence of opposite/minority polarities within the strong majority-polarity plume base.

Our prior studies of jets provide important insights into the physical conditions responsible for jetlets. Transient bipoles frequently emerge within coronal holes, yielding weak minority-polarity regions embedded within the majority polarity flux; these regions are typically observed as coronal bright points (BPs). Jets often are detected all over the CH, wherever a BP exists, and not limited to the vicinity of plumes \citep{kumar2018,kumar2019a,kumar2019b}. Tiny filaments are also observed in most, if not all, BPs that become jet sources \citep{sterling2015,kumar2019a}. Episodic energy releases occur throughout the existence of a typical coronal BP, marked by collimated outflows from and brightenings within the BP. The jet activity stops once the minority polarity decays completely and the BP disappears. 
The key point is that minority polarity surrounded by the background field forms a fan-spine/null-point topology, which is a prime location for current-sheet formation and magnetic reconnection leading to a wide range of solar eruptions \citep{antiochos1998,wyper2017}. In particular, we determined that coronal jets can be generated by a double reconnection process --- the breakout mechanism --- that creates both collimated outflows from and brightenings within the closed flux system. 

Based on the observations reported here, we propose a model for plumes associated with jetlets (Figure \ref{fig14}). Minuscule bipoles that emerge near and within the majority-polarity plume base form fan-spine topologies, with a dome width of $\approx$2-3$\arcsec$ containing a circular/elliptical polarity inversion line. These bipoles can emerge with preexisting free magnetic energy, or build up free energy at the polarity inversion line through local photospheric motions. 
Random emergence and/or twisting of these bipoles in and around the dominant polarity thus produces jetlet sources throughout the plage. The jetlets stop once the majority polarity decays sufficiently, associated with the disappearance of the plume.  The concurrence between the jetlet periodicity and the dominant p-mode period suggests a close connection between the photospheric motions and the repetitive nature of the jetlets, although the actual mechanism is unclear. We propose two possibilities: periodically driven reconnection or periodic modulation of the reconnection outflows.

The cusp-shaped structures underlying the jetlets 
suggest that they are miniature versions of larger jets from bright points (which are similar null-point topologies). By analogy to our simulations of jets, then, the jetlets also are driven by reconnection, specifically by either the resistive-kink mechanism \citep{pariat2009,pariat2010,pariat2015,pariat2016,karpen2017} or the breakout mechanism \citep{wyper2016,wyper2017,wyper2018,wyper2019}. The key difference between these models is that the breakout process involves the formation and eruption of a sheared filament channel within the source, whereas the broader footpoint displacements in the resistive-kink model do not form a filament channel. In both cases we expect a narrow bright ribbon at the base of the fan while fast interchange reconnection is occurring. Only the breakout model predicts bright loops concentrated across the polarity inversion line inside the dome, however, analogous to the flare arcades seen in jets \citep{kumar2018,kumar2019a}. Current observations of jetlets are too coarsely resolved to distinguish between such internal arcades and broader illumination of the separatix dome itself and the closed flux inside it, which could result from a resistive-kink eruption.  As in our jet simulations, we interpret the propagating bright features 
as jet material. 

This study extends the universality of the breakout model for solar eruptions \citep{wyper2017} to the smallest spatial scales observed by AIA and IRIS. A new feature in the jetlets is the 3-5 minute periodicity, however, which could be explained in two ways: (1) reconnection at the null could be triggered by 3-5 minute p-mode oscillations, or (2) p-mode surface motions could push the jetlet sources up and down, generating MHD waves that travel along the reconnection outflow. Spectroscopic observations with high sensitivity and cadence would aid in resolving this issue. 

The detection of quasiperiodic jetlets and energy release at the base of plumes is extremely important for understanding coronal heating and the birth of the nascent solar wind. The plasma temperature in jetlets is $\approx$1-2 MK, while their speeds are of order 100 km s$^{-1}$. The thermal and kinetic energies of most jetlets are $\approx$10$^{24}$ ergs, equal to the energy of a nanoflare \citep{parker1988}. The resulting upper limit on the mass-loss rate ($\approx$1.2$\times$10$^{12}$ g s$^{-1}$) is roughly comparable to the mass-loss rate in the wind at solar minimum. Therefore, the jetlets observed in our study had sufficient mass flux to contribute substantially to the solar wind. Every increase in spatial and temporal resolution has revealed smaller and smaller jetlet-like transient, bright features throughout the solar atmosphere \citep{berghmans2021}.  We expect to observe similar quasiperiodic reconnection at the base of active-region fan loops rooted in plage regions, which requires further investigation. Therefore our insights into jetlets likely extend to quiet-Sun and active-region heating and dynamics.

It is hard to see tiny jets in IRIS disk observations except as multiple brightenings at the base with faint outflow, so they are more easily detected in IRIS and other limb observations. On the other hand, high-resolution H$\alpha$ disk observations are required to resolve cool jetlet structures. Coordinated observations of plume bases with existing instruments (IRIS, GST, Hinode/SOT Ca II) will definitely help to resolve these jetlets.  These instruments all have limited fields of view, however, so coordinated campaigns are crucial to obtain optimal temperature coverage of the same events.

The observational signatures of these frequent, quasiperiodic energy releases may be detected by Parker Solar Probe during its closest approach to the Sun ($\approx$10 R$_\odot$). Several mechanisms (spicules, jetlets, jets) have been proposed as the sources of recently detected PSP switchbacks in the solar wind \citep{sterling2020b}. Another observational signature could be a series of type-III radio bursts excited by accelerated electron beams along the plume, released via repetitive interchange reconnection, which also could produce brightenings detected by the SPICE and EUI instruments on Solar Orbiter (SolO; \citealt{muller2013}). The weak minority polarities within the unipolar plage region in high-resolution magnetograms could be measured by the SolO/PHI instrument. In addition, high-resolution DKIST \citep{tritschler2015} observations (H$\alpha$ and magnetograms) are poised to build a comprehensive picture of jetlets from the photosphere to the corona in the near future. At present, MHD simulations have been focused on single jets, so no MHD model has taken into account quasiperiodic energy release at multiple locations at the base of a plume. New 3D MHD models of coronal heating and generation of the solar wind in coronal plumes should be guided by these exciting observations.

\begin{acknowledgments}
We are grateful to the referee for insightful comments that
have improved this paper. PK thanks Mark Cheung for discussions of DEM analysis. SDO is a mission for NASA's Living With a Star (LWS) program. IRIS is a NASA Small Explorer mission developed and operated by LMSAL with mission operations executed at NASA Ames Research Center and major contributions to downlink communications funded by ESA and the Norwegian Space Centre. 
This research was supported by NASA’s Heliophysics Supporting Research (now part of an Internal Scientist Funding Model work
package) and Guest Investigator (\#80NSSC20K0265) programs. The work of VMU was partly supported through cooperative agreements NNG11PL10A and 80NSSC21M0180 between NASA Goddard Space Flight Center and the Catholic University of America. Magnetic-field extrapolations were visualized with VAPOR (www.vapor.ucar.edu), a product of the Computational Information Systems Laboratory at the National Center for Atmospheric Research. Wavelet software was provided by C. Torrence and G. Compo, and is available at http://paos.colorado.edu/research/wavelets/.
\end{acknowledgments}

\bibliographystyle{aasjournal}
\bibliography{reference.bib}

\begin{thebibliography}{}
\expandafter\ifx\csname natexlab\endcsname\relax\def\natexlab#1{#1}\fi
\providecommand{\url}[1]{\href{#1}{#1}}
\providecommand{\dodoi}[1]{doi:~\href{http://doi.org/#1}{\nolinkurl{#1}}}
\providecommand{\doeprint}[1]{\href{http://ascl.net/#1}{\nolinkurl{http://ascl.net/#1}}}
\providecommand{\doarXiv}[1]{\href{https://arxiv.org/abs/#1}{\nolinkurl{https://arxiv.org/abs/#1}}}

\bibitem[{{Abramenko} \& {Yurchyshyn}(2020)}]{abramenko2020}
{Abramenko}, V.~I., \& {Yurchyshyn}, V.~B. 2020, \mnras, 497, 5405,
  \dodoi{10.1093/mnras/staa2427}

\bibitem[{{Antiochos}(1998)}]{antiochos1998}
{Antiochos}, S.~K. 1998, Astrophys. J. Lett., 502, L181, \dodoi{10.1086/311507}

\bibitem[{{Bale} {et~al.}(2019){Bale}, {Badman}, {Bonnell}, {Bowen}, {Burgess},
  {Case}, {Cattell}, {Chandran}, {Chaston}, {Chen}, {Drake}, {de Wit},
  {Eastwood}, {Ergun}, {Farrell}, {Fong}, {Goetz}, {Goldstein}, {Goodrich},
  {Harvey}, {Horbury}, {Howes}, {Kasper}, {Kellogg}, {Klimchuk}, {Korreck},
  {Krasnoselskikh}, {Krucker}, {Laker}, {Larson}, {MacDowall}, {Maksimovic},
  {Malaspina}, {Martinez-Oliveros}, {McComas}, {Meyer-Vernet}, {Moncuquet},
  {Mozer}, {Phan}, {Pulupa}, {Raouafi}, {Salem}, {Stansby}, {Stevens}, {Szabo},
  {Velli}, {Woolley}, \& {Wygant}}]{bale2019}
{Bale}, S.~D., {Badman}, S.~T., {Bonnell}, J.~W., {et~al.} 2019, \nat, 576,
  237, \dodoi{10.1038/s41586-019-1818-7}

\bibitem[{{Bale} {et~al.}(2021){Bale}, {Horbury}, {Velli}, {Desai}, {Halekas},
  {McManus}, {Panasenco}, {Badman}, {Bowen}, {Chandran}, {Drake}, {Kasper},
  {Laker}, {Mallet}, {Matteini}, {Phan}, {Raouafi}, {Squire}, {Woodham}, \&
  {Woolley}}]{bale2021}
{Bale}, S.~D., {Horbury}, T.~S., {Velli}, M., {et~al.} 2021, \apj, 923, 174,
  \dodoi{10.3847/1538-4357/ac2d8c}

\bibitem[{{Banerjee} \& {Krishna Prasad}(2016)}]{banerjee2016}
{Banerjee}, D., \& {Krishna Prasad}, S. 2016, Washington DC American
  Geophysical Union Geophysical Monograph Series, 216, 419,
  \dodoi{10.1002/9781119055006.ch24}

\bibitem[{{Banerjee} {et~al.}(2021){Banerjee}, {Krishna Prasad}, {Pant},
  {McLaughlin}, {Antolin}, {Magyar}, {Ofman}, {Tian}, {Van Doorsselaere}, {De
  Moortel}, \& {Wang}}]{banerjee2021}
{Banerjee}, D., {Krishna Prasad}, S., {Pant}, V., {et~al.} 2021, \ssr, 217, 76,
  \dodoi{10.1007/s11214-021-00849-0}

\bibitem[{{Berghmans} \& {Clette}(1999)}]{berghmans1999}
{Berghmans}, D., \& {Clette}, F. 1999, \solphys, 186, 207,
  \dodoi{10.1023/A:1005189508371}

\bibitem[{{Berghmans} {et~al.}(2021){Berghmans}, {Auch{\`e}re}, {Long},
  {Soubri{\'e}}, {Mierla}, {Zhukov}, {Sch{\"u}hle}, {Antolin}, {Harra},
  {Parenti}, {Podladchikova}, {Aznar Cuadrado}, {Buchlin}, {Dolla}, {Verbeeck},
  {Gissot}, {Teriaca}, {Haberreiter}, {Katsiyannis}, {Rodriguez}, {Kraaikamp},
  {Smith}, {Stegen}, {Rochus}, {Halain}, {Jacques}, {Thompson}, \&
  {Inhester}}]{berghmans2021}
{Berghmans}, D., {Auch{\`e}re}, F., {Long}, D.~M., {et~al.} 2021, \aap, 656,
  L4, \dodoi{10.1051/0004-6361/202140380}

\bibitem[{{Bigelow}(1891)}]{bigelow1891}
{Bigelow}, F.~H. 1891, The Observatory, 14, 50

\bibitem[{{Chen} \& {Priest}(2006)}]{chen2006}
{Chen}, P.~F., \& {Priest}, E.~R. 2006, Sol. Phys., 238, 313,
  \dodoi{10.1007/s11207-006-0215-1}

\bibitem[{{Cheung} {et~al.}(2015){Cheung}, {Boerner}, {Schrijver}, {Testa},
  {Chen}, {Peter}, \& {Malanushenko}}]{cheung2015}
{Cheung}, M. C.~M., {Boerner}, P., {Schrijver}, C.~J., {et~al.} 2015,
  Astrophys. J., 807, 143, \dodoi{10.1088/0004-637X/807/2/143}

\bibitem[{{Cho} {et~al.}(2020){Cho}, {Nakariakov}, {Moon}, {Lee}, {Yu}, {Cho},
  {Yurchyshyn}, \& {Lee}}]{cho2020}
{Cho}, I.-H., {Nakariakov}, V.~M., {Moon}, Y.-J., {et~al.} 2020, Astrophys. J.
  Lett., 900, L19, \dodoi{10.3847/2041-8213/abb020}

\bibitem[{{Cho} {et~al.}(2021){Cho}, {Cho}, {Madjarska}, {Nakariakov}, {Yang},
  {Choi}, {Lim}, {Lee}, {Seough}, {Lee}, \& {Kim}}]{cho2021}
{Cho}, K.-S., {Cho}, I.-H., {Madjarska}, M.~S., {et~al.} 2021, \apj, 909, 202,
  \dodoi{10.3847/1538-4357/abdfd5}

\bibitem[{{De Pontieu} {et~al.}(2005){De Pontieu}, {Erd{\'e}lyi}, \& {De
  Moortel}}]{depontieu2005}
{De Pontieu}, B., {Erd{\'e}lyi}, R., \& {De Moortel}, I. 2005, Astrophys. J.
  Lett., 624, L61, \dodoi{10.1086/430345}

\bibitem[{{De Pontieu} {et~al.}(2004){De Pontieu}, {Erd{\'e}lyi}, \&
  {James}}]{depontieu2004}
{De Pontieu}, B., {Erd{\'e}lyi}, R., \& {James}, S.~P. 2004, Nature, 430, 536,
  \dodoi{10.1038/nature02749}

\bibitem[{{De Pontieu} {et~al.}(2014){De Pontieu}, {Title}, {Lemen}, {Kushner},
  {Akin}, {Allard}, {Berger}, {Boerner}, {Cheung}, {Chou}, {Drake}, {Duncan},
  {Freeland}, {Heyman}, {Hoffman}, {Hurlburt}, {Lindgren}, {Mathur}, {Rehse},
  {Sabolish}, {Seguin}, {Schrijver}, {Tarbell}, {W{\"u}lser}, {Wolfson},
  {Yanari}, {Mudge}, {Nguyen-Phuc}, {Timmons}, {van Bezooijen}, {Weingrod},
  {Brookner}, {Butcher}, {Dougherty}, {Eder}, {Knagenhjelm}, {Larsen},
  {Mansir}, {Phan}, {Boyle}, {Cheimets}, {DeLuca}, {Golub}, {Gates}, {Hertz},
  {McKillop}, {Park}, {Perry}, {Podgorski}, {Reeves}, {Saar}, {Testa}, {Tian},
  {Weber}, {Dunn}, {Eccles}, {Jaeggli}, {Kankelborg}, {Mashburn}, {Pust},
  {Springer}, {Carvalho}, {Kleint}, {Marmie}, {Mazmanian}, {Pereira}, {Sawyer},
  {Strong}, {Worden}, {Carlsson}, {Hansteen}, {Leenaarts}, {Wiesmann},
  {Aloise}, {Chu}, {Bush}, {Scherrer}, {Brekke}, {Martinez-Sykora}, {Lites},
  {McIntosh}, {Uitenbroek}, {Okamoto}, {Gummin}, {Auker}, {Jerram}, {Pool}, \&
  {Waltham}}]{bart2014}
{De Pontieu}, B., {Title}, A.~M., {Lemen}, J.~R., {et~al.} 2014, Sol. Phys.,
  289, 2733, \dodoi{10.1007/s11207-014-0485-y}

\bibitem[{{DeForest}(2007)}]{deforest2007}
{DeForest}, C.~E. 2007, Astrophys. J., 661, 532, \dodoi{10.1086/515561}

\bibitem[{{DeForest}(2017)}]{deforest2017}
---. 2017, \apj, 838, 155, \dodoi{10.3847/1538-4357/aa67f1}

\bibitem[{{DeForest} \& {Gurman}(1998)}]{deforest1998}
{DeForest}, C.~E., \& {Gurman}, J.~B. 1998, ApJL{Astrophys. J. Lett.}, 501,
  L217, \dodoi{10.1086/311460}

\bibitem[{{DeForest} {et~al.}(1997){DeForest}, {Hoeksema}, {Gurman},
  {Thompson}, {Plunkett}, {Howard}, {Harrison}, \& {Hassler}}]{deforest1997}
{DeForest}, C.~E., {Hoeksema}, J.~T., {Gurman}, J.~B., {et~al.} 1997, Sol.
  Phys., 175, 393, \dodoi{10.1023/A:1004955223306}

\bibitem[{{DeForest} {et~al.}(2001{\natexlab{a}}){DeForest}, {Lamy}, \&
  {Llebaria}}]{deforest2001b}
{DeForest}, C.~E., {Lamy}, P.~L., \& {Llebaria}, A. 2001{\natexlab{a}},
  Astrophys. J., 560, 490, \dodoi{10.1086/322497}

\bibitem[{{DeForest} {et~al.}(2001{\natexlab{b}}){DeForest}, {Plunkett}, \&
  {Andrews}}]{deforest2001a}
{DeForest}, C.~E., {Plunkett}, S.~P., \& {Andrews}, M.~D. 2001{\natexlab{b}},
  Astrophys. J., 546, 569, \dodoi{10.1086/318221}

\bibitem[{{Doyle} {et~al.}(2006){Doyle}, {Popescu}, \& {Taroyan}}]{doyle2006}
{Doyle}, J.~G., {Popescu}, M.~D., \& {Taroyan}, Y. 2006, \aap, 446, 327,
  \dodoi{10.1051/0004-6361:20053826}

\bibitem[{{Fargette} {et~al.}(2021){Fargette}, {Lavraud}, {Rouillard},
  {R{\'e}ville}, {Dudok De Wit}, {Froment}, {Halekas}, {Phan}, {Malaspina},
  {Bale}, {Kasper}, {Louarn}, {Case}, {Korreck}, {Larson}, {Pulupa}, {Stevens},
  {Whittlesey}, \& {Berthomier}}]{Fargette2021}
{Fargette}, N., {Lavraud}, B., {Rouillard}, A.~P., {et~al.} 2021, \apj, 919,
  96, \dodoi{10.3847/1538-4357/ac1112}

\bibitem[{{Fisher} \& {Guhathakurta}(1995)}]{fisher1995}
{Fisher}, R., \& {Guhathakurta}, M. 1995, Astrophys. J. Lett., 447, L139,
  \dodoi{10.1086/309582}

\bibitem[{{Fisk} \& {Kasper}(2020)}]{fisk2020}
{Fisk}, L.~A., \& {Kasper}, J.~C. 2020, \apjl, 894, L4,
  \dodoi{10.3847/2041-8213/ab8acd}

\bibitem[{{Fox} {et~al.}(2016){Fox}, {Velli}, {Bale}, {Decker}, {Driesman},
  {Howard}, {Kasper}, {Kinnison}, {Kusterer}, {Lario}, {Lockwood}, {McComas},
  {Raouafi}, \& {Szabo}}]{fox2016}
{Fox}, N.~J., {Velli}, M.~C., {Bale}, S.~D., {et~al.} 2016, Space Sci. Rev.,
  204, 7, \dodoi{10.1007/s11214-015-0211-6}

\bibitem[{{Gupta} {et~al.}(2012){Gupta}, {Teriaca}, {Marsch}, {Solanki}, \&
  {Banerjee}}]{gupta2012}
{Gupta}, G.~R., {Teriaca}, L., {Marsch}, E., {Solanki}, S.~K., \& {Banerjee},
  D. 2012, Astron. Astroph., 546, A93, \dodoi{10.1051/0004-6361/201219795}

\bibitem[{{Gupta} \& {Tripathi}(2015)}]{gupta2015}
{Gupta}, G.~R., \& {Tripathi}, D. 2015, \apj, 809, 82,
  \dodoi{10.1088/0004-637X/809/1/82}

\bibitem[{{Heggland} {et~al.}(2009){Heggland}, {De Pontieu}, \&
  {Hansteen}}]{Heggland2009}
{Heggland}, L., {De Pontieu}, B., \& {Hansteen}, V.~H. 2009, \apj, 702, 1,
  \dodoi{10.1088/0004-637X/702/1/1}

\bibitem[{{Heggland} {et~al.}(2011){Heggland}, {Hansteen}, {De Pontieu}, \&
  {Carlsson}}]{heggland2011}
{Heggland}, L., {Hansteen}, V.~H., {De Pontieu}, B., \& {Carlsson}, M. 2011,
  \apj, 743, 142, \dodoi{10.1088/0004-637X/743/2/142}

\bibitem[{Horbury {et~al.}(2020)Horbury, Woolley, Laker, Matteini, Eastwood,
  Bale, Velli, Chandran, Phan, Raouafi, Goetz, Harvey, Pulupa, Klein, Wit,
  Kasper, Korreck, Case, Stevens, Whittlesey, Larson, MacDowall, Malaspina, \&
  Livi}]{horbury2020}
Horbury, T.~S., Woolley, T., Laker, R., {et~al.} 2020, The Astrophysical
  Journal Supplement Series, 246, 45, \dodoi{10.3847/1538-4365/ab5b15}

\bibitem[{{Jiao} {et~al.}(2015){Jiao}, {Xia}, {Li}, {Huang}, {Li},
  {Chandrashekhar}, {Mou}, \& {Fu}}]{jiao2015}
{Jiao}, F., {Xia}, L., {Li}, B., {et~al.} 2015, \apjl, 809, L17,
  \dodoi{10.1088/2041-8205/809/1/L17}

\bibitem[{{Karpen} {et~al.}(2017){Karpen}, {DeVore}, {Antiochos}, \&
  {Pariat}}]{karpen2017}
{Karpen}, J.~T., {DeVore}, C.~R., {Antiochos}, S.~K., \& {Pariat}, E. 2017,
  \apj, 834, 62, \dodoi{10.3847/1538-4357/834/1/62}

\bibitem[{{Kasper} {et~al.}(2019){Kasper}, {Bale}, {Belcher}, {Berthomier},
  {Case}, {Chandran}, {Curtis}, {Gallagher}, {Gary}, {Golub}, {Halekas}, {Ho},
  {Horbury}, {Hu}, {Huang}, {Klein}, {Korreck}, {Larson}, {Livi}, {Maruca},
  {Lavraud}, {Louarn}, {Maksimovic}, {Martinovic}, {McGinnis}, {Pogorelov},
  {Richardson}, {Skoug}, {Steinberg}, {Stevens}, {Szabo}, {Velli},
  {Whittlesey}, {Wright}, {Zank}, {MacDowall}, {McComas}, {McNutt}, {Pulupa},
  {Raouafi}, \& {Schwadron}}]{kasper2019}
{Kasper}, J.~C., {Bale}, S.~D., {Belcher}, J.~W., {et~al.} 2019, Nature, 576,
  228, \dodoi{10.1038/s41586-019-1813-z}

\bibitem[{{Keiling} {et~al.}(2012){Keiling}, {Shiokawa}, {Uritsky}, {Sergeev},
  {Zesta}, {Kepko}, \& {{\O}stgaard}}]{keiling2012}
{Keiling}, A., {Shiokawa}, K., {Uritsky}, V., {et~al.} 2012, Washington DC
  American Geophysical Union Geophysical Monograph Series, 197, 317,
  \dodoi{10.1029/2012GM001231}

\bibitem[{Khomenko \& Cally(2019)}]{khomenko2019}
Khomenko, E., \& Cally, P.~S. 2019, Astrophys. J., 883, 179,
  \dodoi{10.3847/1538-4357/ab3d28}

\bibitem[{{Klimchuk}(2006)}]{klimchuk2006}
{Klimchuk}, J.~A. 2006, Sol. Phys., 234, 41, \dodoi{10.1007/s11207-006-0055-z}

\bibitem[{{Klimchuk}(2015)}]{klimchuk2015}
---. 2015, Philosophical Transactions of the Royal Society of London Series A,
  373, 20140256, \dodoi{10.1098/rsta.2014.0256}

\bibitem[{{Krishna Prasad} {et~al.}(2011){Krishna Prasad}, {Banerjee}, \&
  {Gupta}}]{krishna2011}
{Krishna Prasad}, S., {Banerjee}, D., \& {Gupta}, G.~R. 2011, \aap, 528, L4,
  \dodoi{10.1051/0004-6361/201016405}

\bibitem[{{Krishna Prasad} {et~al.}(2014){Krishna Prasad}, {Banerjee}, \& {Van
  Doorsselaere}}]{Krishna2014}
{Krishna Prasad}, S., {Banerjee}, D., \& {Van Doorsselaere}, T. 2014, \apj,
  789, 118, \dodoi{10.1088/0004-637X/789/2/118}

\bibitem[{{Kumar} {et~al.}(2019{\natexlab{a}}){Kumar}, {Karpen}, {Antiochos},
  {Wyper}, \& {DeVore}}]{kumar2019b}
{Kumar}, P., {Karpen}, J.~T., {Antiochos}, S.~K., {Wyper}, P.~F., \& {DeVore},
  C.~R. 2019{\natexlab{a}}, Astrophys. J. Lett., 885, L15,
  \dodoi{10.3847/2041-8213/ab45f9}

\bibitem[{{Kumar} {et~al.}(2018){Kumar}, {Karpen}, {Antiochos}, {Wyper},
  {DeVore}, \& {DeForest}}]{kumar2018}
{Kumar}, P., {Karpen}, J.~T., {Antiochos}, S.~K., {et~al.} 2018, Astrophys. J.,
  854, 155, \dodoi{10.3847/1538-4357/aaab4f}

\bibitem[{{Kumar} {et~al.}(2019{\natexlab{b}}){Kumar}, {Karpen}, {Antiochos},
  {Wyper}, {DeVore}, \& {DeForest}}]{kumar2019a}
---. 2019{\natexlab{b}}, Astrophys. J., 873, 93,
  \dodoi{10.3847/1538-4357/ab04af}

\bibitem[{{Kumar} {et~al.}(2021){Kumar}, {Karpen}, {Antiochos}, {Wyper},
  {DeVore}, \& {Lynch}}]{kumar2021}
---. 2021, Astrophys. J., 907, 41, \dodoi{10.3847/1538-4357/abca8b}

\bibitem[{{Kumar} {et~al.}(2015){Kumar}, {Nakariakov}, \& {Cho}}]{kumar2015}
{Kumar}, P., {Nakariakov}, V.~M., \& {Cho}, K.-S. 2015, Astrophys. J., 804, 4,
  \dodoi{10.1088/0004-637X/804/1/4}

\bibitem[{{Kumar} {et~al.}(2016){Kumar}, {Nakariakov}, \& {Cho}}]{kumar2016}
---. 2016, Astrophys. J., 822, 7, \dodoi{10.3847/0004-637X/822/1/7}

\bibitem[{{Lee} {et~al.}(2020){Lee}, {Karpen}, {Liu}, \& {Wang}}]{lee2020}
{Lee}, J., {Karpen}, J.~T., {Liu}, C., \& {Wang}, H. 2020, Astrophys. J., 893,
  158, \dodoi{10.3847/1538-4357/ab80c4}

\bibitem[{{Lemen} {et~al.}(2012){Lemen}, {Title}, {Akin}, {Boerner}, {Chou},
  {Drake}, {Duncan}, {Edwards}, {Friedlaender}, {Heyman}, {Hurlburt}, {Katz},
  {Kushner}, {Levay}, {Lindgren}, {Mathur}, {McFeaters}, {Mitchell}, {Rehse},
  {Schrijver}, {Springer}, {Stern}, {Tarbell}, {Wuelser}, {Wolfson}, {Yanari},
  {Bookbinder}, {Cheimets}, {Caldwell}, {Deluca}, {Gates}, {Golub}, {Park},
  {Podgorski}, {Bush}, {Scherrer}, {Gummin}, {Smith}, {Auker}, {Jerram},
  {Pool}, {Soufli}, {Windt}, {Beardsley}, {Clapp}, {Lang}, \&
  {Waltham}}]{lemen2012}
{Lemen}, J.~R., {Title}, A.~M., {Akin}, D.~J., {et~al.} 2012, Sol. Phys., 275,
  17, \dodoi{10.1007/s11207-011-9776-8}

\bibitem[{{Masson} {et~al.}(2009){Masson}, {Pariat}, {Aulanier}, \&
  {Schrijver}}]{masson2009}
{Masson}, S., {Pariat}, E., {Aulanier}, G., \& {Schrijver}, C.~J. 2009,
  Astrophys. J., 700, 559, \dodoi{10.1088/0004-637X/700/1/559}

\bibitem[{{McIntosh} {et~al.}(2010){McIntosh}, {Innes}, {de Pontieu}, \&
  {Leamon}}]{mcintosh2010}
{McIntosh}, S.~W., {Innes}, D.~E., {de Pontieu}, B., \& {Leamon}, R.~J. 2010,
  Astron. Astroph., 510, L2, \dodoi{10.1051/0004-6361/200913699}

\bibitem[{Morton {et~al.}(2015)Morton, Tomczyk, \& Pinto}]{morton2015}
Morton, R.~J., Tomczyk, S., \& Pinto, R. 2015, Nature Communications, 6, 7813,
  \dodoi{10.1038/ncomms8813}

\bibitem[{{M{\"u}ller} {et~al.}(2013){M{\"u}ller}, {Marsden}, {St. Cyr}, \&
  {Gilbert}}]{muller2013}
{M{\"u}ller}, D., {Marsden}, R.~G., {St. Cyr}, O.~C., \& {Gilbert}, H.~R. 2013,
  Sol. Phys., 285, 25, \dodoi{10.1007/s11207-012-0085-7}

\bibitem[{{Nakagawa} \& {Raadu}(1972)}]{nakagawa1972}
{Nakagawa}, Y., \& {Raadu}, M.~A. 1972, \solphys, 25, 127,
  \dodoi{10.1007/BF00155751}

\bibitem[{{Neugebauer}(2012)}]{neugebauer2012}
{Neugebauer}, M. 2012, \apj, 750, 50, \dodoi{10.1088/0004-637X/750/1/50}

\bibitem[{{Neugebauer} {et~al.}(1995){Neugebauer}, {Goldstein}, {McComas},
  {Suess}, \& {Balogh}}]{neugebauer1995}
{Neugebauer}, M., {Goldstein}, B.~E., {McComas}, D.~J., {Suess}, S.~T., \&
  {Balogh}, A. 1995, \jgr, 100, 23389, \dodoi{10.1029/95JA02723}

\bibitem[{{Neugebauer} \& {Sterling}(2021)}]{neugebauer2021}
{Neugebauer}, M., \& {Sterling}, A.~C. 2021, \apjl, 920, L31,
  \dodoi{10.3847/2041-8213/ac2945}

\bibitem[{{Ning} {et~al.}(2004){Ning}, {Innes}, \& {Solanki}}]{ning2004}
{Ning}, Z., {Innes}, D.~E., \& {Solanki}, S.~K. 2004, \aap, 419, 1141,
  \dodoi{10.1051/0004-6361:20034499}

\bibitem[{{Nita} {et~al.}(2015){Nita}, {Fleishman}, {Kuznetsov}, {Kontar}, \&
  {Gary}}]{nita2015}
{Nita}, G.~M., {Fleishman}, G.~D., {Kuznetsov}, A.~A., {Kontar}, E.~P., \&
  {Gary}, D.~E. 2015, \apj, 799, 236, \dodoi{10.1088/0004-637X/799/2/236}

\bibitem[{{Ofman} {et~al.}(1999){Ofman}, {Nakariakov}, \&
  {DeForest}}]{ofman1999}
{Ofman}, L., {Nakariakov}, V.~M., \& {DeForest}, C.~E. 1999, Astrophys. J.,
  514, 441, \dodoi{10.1086/306944}

\bibitem[{{Ofman} {et~al.}(2012){Ofman}, {Wang}, \& {Davila}}]{ofman2012}
{Ofman}, L., {Wang}, T.~J., \& {Davila}, J.~M. 2012, \apj, 754, 111,
  \dodoi{10.1088/0004-637X/754/2/111}

\bibitem[{{Panesar} {et~al.}(2018){Panesar}, {Sterling}, {Moore}, {Tiwari}, {De
  Pontieu}, \& {Norton}}]{panesar2018}
{Panesar}, N.~K., {Sterling}, A.~C., {Moore}, R.~L., {et~al.} 2018, Astrophys.
  J. Lett., 868, L27, \dodoi{10.3847/2041-8213/aaef37}

\bibitem[{{Pant} {et~al.}(2015){Pant}, {Dolla}, {Mazumder}, {Banerjee},
  {Krishna Prasad}, \& {Panditi}}]{pant2015}
{Pant}, V., {Dolla}, L., {Mazumder}, R., {et~al.} 2015, \apj, 807, 71,
  \dodoi{10.1088/0004-637X/807/1/71}

\bibitem[{{Pariat} {et~al.}(2009){Pariat}, {Antiochos}, \&
  {DeVore}}]{pariat2009}
{Pariat}, E., {Antiochos}, S.~K., \& {DeVore}, C.~R. 2009, Astrophys. J., 691,
  61, \dodoi{10.1088/0004-637X/691/1/61}

\bibitem[{{Pariat} {et~al.}(2010){Pariat}, {Antiochos}, \&
  {DeVore}}]{pariat2010}
---. 2010, Astrophys. J., 714, 1762, \dodoi{10.1088/0004-637X/714/2/1762}

\bibitem[{{Pariat} {et~al.}(2015){Pariat}, {Dalmasse}, {DeVore}, {Antiochos},
  \& {Karpen}}]{pariat2015}
{Pariat}, E., {Dalmasse}, K., {DeVore}, C.~R., {Antiochos}, S.~K., \& {Karpen},
  J.~T. 2015, Astronomy and Astrophysics, 573, A130,
  \dodoi{10.1051/0004-6361/201424209}

\bibitem[{{Pariat} {et~al.}(2016){Pariat}, {Dalmasse}, {DeVore}, {Antiochos},
  \& {Karpen}}]{pariat2016}
---. 2016, Astronomy and Astrophysics, 596, A36,
  \dodoi{10.1051/0004-6361/201629109}

\bibitem[{{Parker}(1988)}]{parker1988}
{Parker}, E.~N. 1988, Astrophys. J., 330, 474, \dodoi{10.1086/166485}

\bibitem[{{Petralia} {et~al.}(2014){Petralia}, {Reale}, {Orlando}, \&
  {Klimchuk}}]{petralia2014}
{Petralia}, A., {Reale}, F., {Orlando}, S., \& {Klimchuk}, J.~A. 2014, \aap,
  567, A70, \dodoi{10.1051/0004-6361/201323012}

\bibitem[{{Poletto}(2015)}]{poletto2015}
{Poletto}, G. 2015, Living Reviews in Solar Physics, 12, 7,
  \dodoi{10.1007/lrsp-2015-7}

\bibitem[{{Raouafi} {et~al.}(2007){Raouafi}, {Harvey}, \&
  {Solanki}}]{raouafi2007}
{Raouafi}, N.~E., {Harvey}, J.~W., \& {Solanki}, S.~K. 2007, Astrophys. J.,
  658, 643, \dodoi{10.1086/510286}

\bibitem[{{Raouafi} {et~al.}(2008){Raouafi}, {Petrie}, {Norton}, {Henney}, \&
  {Solanki}}]{raouafi2008}
{Raouafi}, N.~E., {Petrie}, G.~J.~D., {Norton}, A.~A., {Henney}, C.~J., \&
  {Solanki}, S.~K. 2008, Astrophys. J. Lett., 682, L137, \dodoi{10.1086/591125}

\bibitem[{{Raouafi} \& {Stenborg}(2014)}]{raouafi2014}
{Raouafi}, N.~E., \& {Stenborg}, G. 2014, \apj, 787, 118,
  \dodoi{10.1088/0004-637X/787/2/118}

\bibitem[{{R{\'e}gnier} {et~al.}(2014){R{\'e}gnier}, {Alexander}, {Walsh},
  {Winebarger}, {Cirtain}, {Golub}, {Korreck}, {Mitchell}, {Platt}, {Weber},
  {De Pontieu}, {Title}, {Kobayashi}, {Kuzin}, \& {DeForest}}]{regnier2014}
{R{\'e}gnier}, S., {Alexander}, C.~E., {Walsh}, R.~W., {et~al.} 2014, \apj,
  784, 134, \dodoi{10.1088/0004-637X/784/2/134}

\bibitem[{{Roberts} {et~al.}(2018){Roberts}, {Uritsky}, {DeVore}, \&
  {Karpen}}]{roberts2018}
{Roberts}, M.~A., {Uritsky}, V.~M., {DeVore}, C.~R., \& {Karpen}, J.~T. 2018,
  \apj, 866, 14, \dodoi{10.3847/1538-4357/aadb41}

\bibitem[{{Saito}(1958)}]{saito1958}
{Saito}, K. 1958, PASJ, 10, 49

\bibitem[{{Sakao} {et~al.}(2007){Sakao}, {Kano}, {Narukage}, {Kotoku}, {Bando},
  {DeLuca}, {Lundquist}, {Tsuneta}, {Harra}, {Katsukawa}, {Kubo}, {Hara},
  {Matsuzaki}, {Shimojo}, {Bookbinder}, {Golub}, {Korreck}, {Su}, {Shibasaki},
  {Shimizu}, \& {Nakatani}}]{sakao2007}
{Sakao}, T., {Kano}, R., {Narukage}, N., {et~al.} 2007, Science, 318, 1585,
  \dodoi{10.1126/science.1147292}

\bibitem[{{Samanta} {et~al.}(2015){Samanta}, {Pant}, \&
  {Banerjee}}]{samanta2015}
{Samanta}, T., {Pant}, V., \& {Banerjee}, D. 2015, \apjl, 815, L16,
  \dodoi{10.1088/2041-8205/815/1/L16}

\bibitem[{{Samanta} {et~al.}(2019){Samanta}, {Tian}, {Yurchyshyn}, {Peter},
  {Cao}, {Sterling}, {Erd{\'e}lyi}, {Ahn}, {Feng}, {Utz}, {Banerjee}, \&
  {Chen}}]{samanta2019}
{Samanta}, T., {Tian}, H., {Yurchyshyn}, V., {et~al.} 2019, Science, 366, 890,
  \dodoi{10.1126/science.aaw2796}

\bibitem[{{Schou} {et~al.}(2012){Schou}, {Scherrer}, {Bush}, {Wachter},
  {Couvidat}, {Rabello-Soares}, {Bogart}, {Hoeksema}, {Liu}, {Duvall}, {Akin},
  {Allard}, {Miles}, {Rairden}, {Shine}, {Tarbell}, {Title}, {Wolfson},
  {Elmore}, {Norton}, \& {Tomczyk}}]{schou2012}
{Schou}, J., {Scherrer}, P.~H., {Bush}, R.~I., {et~al.} 2012, Sol. Phys., 275,
  229, \dodoi{10.1007/s11207-011-9842-2}

\bibitem[{{Sterling} \& {Moore}(2020)}]{sterling2020a}
{Sterling}, A.~C., \& {Moore}, R.~L. 2020, \apjl, 896, L18,
  \dodoi{10.3847/2041-8213/ab96be}

\bibitem[{{Sterling} {et~al.}(2015){Sterling}, {Moore}, {Falconer}, \&
  {Adams}}]{sterling2015}
{Sterling}, A.~C., {Moore}, R.~L., {Falconer}, D.~A., \& {Adams}, M. 2015,
  Nature, 523, 437, \dodoi{10.1038/nature14556}

\bibitem[{{Sterling} {et~al.}(2020){Sterling}, {Moore}, {Panesar}, \&
  {Samanta}}]{sterling2020b}
{Sterling}, A.~C., {Moore}, R.~L., {Panesar}, N.~K., \& {Samanta}, T. 2020, in
  Journal of Physics Conference Series, Vol. 1620, Journal of Physics
  Conference Series, 012020, \dodoi{10.1088/1742-6596/1620/1/012020}

\bibitem[{{Sych} {et~al.}(2009){Sych}, {Nakariakov}, {Karlicky}, \&
  {Anfinogentov}}]{sych2009}
{Sych}, R., {Nakariakov}, V.~M., {Karlicky}, M., \& {Anfinogentov}, S. 2009,
  \aap, 505, 791, \dodoi{10.1051/0004-6361/200912132}

\bibitem[{{Thieme} {et~al.}(1990){Thieme}, {Marsch}, \& {Schwenn}}]{thieme1990}
{Thieme}, K.~M., {Marsch}, E., \& {Schwenn}, R. 1990, Annales Geophysicae, 8,
  713

\bibitem[{{Thurgood} {et~al.}(2019){Thurgood}, {Pontin}, \&
  {McLaughlin}}]{thurgood2019}
{Thurgood}, J.~O., {Pontin}, D.~I., \& {McLaughlin}, J.~A. 2019, Astron.
  Astroph., 621, A106, \dodoi{10.1051/0004-6361/201834369}

\bibitem[{{Tian} {et~al.}(2011){Tian}, {McIntosh}, {Habbal}, \&
  {He}}]{tian2011}
{Tian}, H., {McIntosh}, S.~W., {Habbal}, S.~R., \& {He}, J. 2011, Astrophys.
  J., 736, 130, \dodoi{10.1088/0004-637X/736/2/130}

\bibitem[{{Tian} {et~al.}(2014){Tian}, {DeLuca}, {Cranmer}, {De Pontieu},
  {Peter}, {Mart{\'\i}nez-Sykora}, {Golub}, {McKillop}, {Reeves}, {Miralles},
  {McCauley}, {Saar}, {Testa}, {Weber}, {Murphy}, {Lemen}, {Title}, {Boerner},
  {Hurlburt}, {Tarbell}, {Wuelser}, {Kleint}, {Kankelborg}, {Jaeggli},
  {Carlsson}, {Hansteen}, \& {McIntosh}}]{tian2014}
{Tian}, H., {DeLuca}, E.~E., {Cranmer}, S.~R., {et~al.} 2014, Science, 346,
  1255711, \dodoi{10.1126/science.1255711}

\bibitem[{Tomczyk \& McIntosh(2009)}]{tomczyk2009}
Tomczyk, S., \& McIntosh, S.~W. 2009, Astrophys. J., 697, 1384,
  \dodoi{10.1088/0004-637x/697/2/1384}

\bibitem[{{Torrence} \& {Compo}(1998)}]{torrence1998}
{Torrence}, C., \& {Compo}, G.~P. 1998, Bulletin of the American Meteorological
  Society, 79, 61, \dodoi{10.1175/1520-0477(1998)079<0061:APGTWA>2.0.CO;2}

\bibitem[{{Tritschler} {et~al.}(2015){Tritschler}, {Rimmele}, {Berukoff},
  {Casini}, {Craig}, {Elmore}, {Hubbard}, {Kuhn}, {Lin}, {McMullin}, {Reardon},
  {Schmidt}, {Warner}, \& {Woger}}]{tritschler2015}
{Tritschler}, A., {Rimmele}, T.~R., {Berukoff}, S., {et~al.} 2015, in Cambridge
  Workshop on Cool Stars, Stellar Systems, and the Sun, Vol.~18, 18th Cambridge
  Workshop on Cool Stars, Stellar Systems, and the Sun, 933--944

\bibitem[{{Tu} {et~al.}(2005){Tu}, {Zhou}, {Marsch}, {Xia}, {Zhao}, {Wang}, \&
  {Wilhelm}}]{tu2005}
{Tu}, C.-Y., {Zhou}, C., {Marsch}, E., {et~al.} 2005, Science, 308, 519,
  \dodoi{10.1126/science.1109447}

\bibitem[{{Uritsky} {et~al.}(2013){Uritsky}, {Davila}, {Viall}, \&
  {Ofman}}]{uritsky2013}
{Uritsky}, V.~M., {Davila}, J.~M., {Viall}, N.~M., \& {Ofman}, L. 2013,
  Astrophys. J., 778, 26, \dodoi{10.1088/0004-637X/778/1/26}

\bibitem[{{Uritsky} {et~al.}(2021){Uritsky}, {DeForest}, {Karpen}, {DeVore},
  {Kumar}, {Raouafi}, \& {Wyper}}]{uritsky2021}
{Uritsky}, V.~M., {DeForest}, C.~E., {Karpen}, J.~T., {et~al.} 2021, Astrophys.
  J., 907, 1, \dodoi{10.3847/1538-4357/abd186}

\bibitem[{{Uritsky} {et~al.}(2009){Uritsky}, {Liang}, {Donovan}, {Spanswick},
  {Knudsen}, {Liu}, {Bonnell}, \& {Glassmeier}}]{uritsky2009}
{Uritsky}, V.~M., {Liang}, J., {Donovan}, E., {et~al.} 2009,
  Geophycal.~Res.~Lett., 36, L21103, \dodoi{10.1029/2009GL040777}

\bibitem[{{Uritsky} {et~al.}(2017){Uritsky}, {Roberts}, {DeVore}, \&
  {Karpen}}]{uritsky2017}
{Uritsky}, V.~M., {Roberts}, M.~A., {DeVore}, C.~R., \& {Karpen}, J.~T. 2017,
  \apj, 837, 123, \dodoi{10.3847/1538-4357/aa5cb9}

\bibitem[{{Wang} \& {Liu}(2012)}]{wang2012}
{Wang}, H., \& {Liu}, C. 2012, Astrophys. J., 760, 101,
  \dodoi{10.1088/0004-637X/760/2/101}

\bibitem[{{Wang} {et~al.}(2013){Wang}, {Ofman}, \& {Davila}}]{wang2013}
{Wang}, T., {Ofman}, L., \& {Davila}, J.~M. 2013, \apjl, 775, L23,
  \dodoi{10.1088/2041-8205/775/1/L23}

\bibitem[{{Wang}(2016)}]{wang2016}
{Wang}, T.~J. 2016, Washington DC American Geophysical Union Geophysical
  Monograph Series, 216, 395, \dodoi{10.1002/9781119055006.ch23}

\bibitem[{{Wang}(1998)}]{wang1998}
{Wang}, Y.~M. 1998, in Astronomical Society of the Pacific Conference Series,
  Vol. 154, Cool Stars, Stellar Systems, and the Sun, ed. R.~A. {Donahue} \&
  J.~A. {Bookbinder}, 131

\bibitem[{{Wang}(2020)}]{wang2020}
{Wang}, Y.~M. 2020, Astrophys. J., 904, 199, \dodoi{10.3847/1538-4357/abbda6}

\bibitem[{{Woo} \& {Habbal}(1997)}]{woo1997}
{Woo}, R., \& {Habbal}, S.~R. 1997, GRL, 24, 1159, \dodoi{10.1029/97GL01156}

\bibitem[{{Wyper} {et~al.}(2017){Wyper}, {Antiochos}, \& {DeVore}}]{wyper2017}
{Wyper}, P.~F., {Antiochos}, S.~K., \& {DeVore}, C.~R. 2017, Nature, 544, 452,
  \dodoi{10.1038/nature22050}

\bibitem[{{Wyper} {et~al.}(2019){Wyper}, {DeVore}, \& {Antiochos}}]{wyper2019}
{Wyper}, P.~F., {DeVore}, C.~R., \& {Antiochos}, S.~K. 2019, MNRAS, 490, 3679,
  \dodoi{10.1093/mnras/stz2674}

\bibitem[{{Wyper} {et~al.}(2018){Wyper}, {DeVore}, {Karpen}, {Antiochos}, \&
  {Yeates}}]{wyper2018}
{Wyper}, P.~F., {DeVore}, C.~R., {Karpen}, J.~T., {Antiochos}, S.~K., \&
  {Yeates}, A.~R. 2018, \apj, 864, 165, \dodoi{10.3847/1538-4357/aad9f7}

\bibitem[{{Wyper} {et~al.}(2016){Wyper}, {DeVore}, {Karpen}, \&
  {Lynch}}]{wyper2016}
{Wyper}, P.~F., {DeVore}, C.~R., {Karpen}, J.~T., \& {Lynch}, B.~J. 2016,
  Astrophys. J., 827, 4, \dodoi{10.3847/0004-637X/827/1/4}

\bibitem[{{Yuan} {et~al.}(2016){Yuan}, {Su}, {Jiao}, \& {Walsh}}]{yuan2016}
{Yuan}, D., {Su}, J., {Jiao}, F., \& {Walsh}, R.~W. 2016, \apjs, 224, 30,
  \dodoi{10.3847/0067-0049/224/2/30}

\bibitem[{{Zank} {et~al.}(2020){Zank}, {Nakanotani}, {Zhao}, {Adhikari}, \&
  {Kasper}}]{zank2020}
{Zank}, G.~P., {Nakanotani}, M., {Zhao}, L.~L., {Adhikari}, L., \& {Kasper}, J.
  2020, \apj, 903, 1, \dodoi{10.3847/1538-4357/abb828}

\end{thebibliography}

{\small

\begin{longtable*}{c c c c c c c}
\caption{Plumes analyzed in this paper} \\
\hline \\
\label{tab1}
Plume   &Date                            &Time     &Location             &Periodicity of footpoint        & Periodicity of      &Speed$^\star$ of  \\
              &                                   & (UT)     & (X,Y in arcsec)  &brightenings in AIA                  & jetlets            & jetlets \\
              &                                   &              &                           & 171/193 \AA~                       & 171/193 \AA~      & (193 \AA)        \\
              &                                   &               &                         &  (min)                                      & (min)                    &\kms      \\
    \hline
 &                                              &                 &                           &                                               &                           &          \\  
P1       &July 3,                        &14:31:36-   &(-400,290)          &3, 5-6,10                                       & $\approx$5       & 106   \\    
           &  2016                         &17:27:24    &                           &                                                &                          &                  \\
            &                                  &                  &                            &                                                &                            &         \\
            
P2       &Oct. 20,                        &11:01:36-   &(-100,350)          &3-5,10                                       & $\approx$5            &65    \\    
           &  2015                         &13:57:24    &                           &                                                &                          &                  \\
           &                                  &                  &                            &                                                &                            &         \\

P3       &March 19,                   &12:56:36-   &(5,370)            & 3-5, 10                                      & $\approx$5                       &110    \\    
           &  2016                         &20:02:48    &                           &                                                &                          &                  \\

           &                                   &                 &                            &                                                &                            &          \\    
 
P4       &Dec. 23,                        &14:01:35-   &(-60,950)          &3-5, 10                                      &$\approx$5                        &130    \\    
           &  2019                         &17:57:23    &                           &                                                &                          &                  \\

           &                                   &                 &                            &                                                &                            &          \\ 
\hline
\end{longtable*}
\small
\noindent
${}^\star$ The observed speed is measured from the TD plots showing jetlets along with footpoint brightenings.\\

}


\end{document}